\documentclass[final,3p,times]{elsarticle}
\usepackage[T1]{fontenc}
\usepackage[utf8]{inputenc}
\usepackage{graphicx}
\usepackage{amssymb}
\usepackage{amsmath}
\usepackage{color}
\usepackage{siunitx}
\usepackage{subcaption}
\usepackage{booktabs}
\usepackage{hyperref}
\usepackage{cleveref}
\usepackage[version=4]{mhchem}
\usepackage{tikz}
\usepackage{pgfplots}
\usepackage{pgfplotstable}

\newcommand{\thetitle}{A multiscale porous--resolved methodology for efficient simulation of heat and fluid transport in complex geometries, with application to electric power transformers}
\journal{Applied Thermal Engineering}

\graphicspath{{./figures/}{./tikz/}}
\usetikzlibrary{arrows.meta}
\usetikzlibrary{plotmarks}
\usepgfplotslibrary{external}
\usepgfplotslibrary{colorbrewer}
\usepgfplotslibrary{polar}
\usepgfplotslibrary{colormaps}
\pgfplotsset{
  scale=1.1,
  compat=newest,
  every axis/.append style={
    height=0.60\textwidth,
    width=0.80\textwidth,
    axis lines=left,
    axis line style={thick, -{Latex[length=5pt,width=5pt]}},
    every tick/.style={color=gray},
    legend style = {
      draw=none,
      fill=black!3,
      font=\footnotesize,
      at={(1, 1.05)},
      anchor=south east,
    },
    legend cell align = {left},
    legend image code/.code={
      \draw[mark repeat=2,mark phase=2]
      plot coordinates {
        (0cm, 0cm)
        (0.2cm, 0cm)
        (0.4cm, 0cm)
      };
    },
    grid style = {line width=.1pt, draw=gray!30},
    major grid style = {line width=.2pt, draw=gray!50},
    font=\small,
    samples=100,
    every axis plot/.append style={very thick},
    set layers=axis on top,
  },
  regression/.style={color=Set1-B},
  p1/.style={},
  p2/.style={densely dotted},
  p3/.style={densely dashed},
  dp1/.style={mark=*, mark options={scale=0.8, solid}},
  dp2/.style={only marks, mark=x, mark options={scale=1.2, solid}},
  dp3/.style={mark=triangle*, mark options={scale=0.9, solid}},
  c21/.style={color=Set1-B},
  c22/.style={color=Set1-E},
  c31/.style={color=Set1-B},
  c32/.style={color=Set1-E},
  c33/.style={color=Set1-C},
  table/col sep = comma,
}
\pgfplotstableset{
  set thousands separator={\,},
  every head row/.style={before row=\toprule,after row=\midrule},
  every last row/.style={after row=\bottomrule}
}

\usepackage[colorinlistoftodos,color=black!30]{todonotes}
\presetkeys{todonotes}{inline}{}
\makeatletter
\renewcommand{\todo}[2][]{%
  \tikzexternaldisable\@todo[#1]{#2}\tikzexternalenable%
}
\makeatother

\biboptions{sort&compress}

\hypersetup{
  pdfauthor = {Ole H.~H.~Meyer, Karl Yngve Lervåg, Åsmund Ervik},
  pdftitle = {\thetitle},
  pdfsubject = {},
  pdfkeywords = {},
  pdfcreator = {LaTeX with hyperref package},
  pdfproducer = {pdflatex},
  bookmarksnumbered,
  pdfstartview={FitH},
  colorlinks=true,
}

\crefname{equation}{Eq.}{Eqs.}
\crefname{section}{Sec.}{Secs.}
\crefname{table}{Tab.}{Tabs.}
\crefname{figure}{Fig.}{Figs.}
\crefname{subfigure}{Fig.}{Figs.}

\newcommand*{\grad}{\ensuremath{\nabla}}
\renewcommand*{\div}{\ensuremath{\nabla\cdot}}
\newcommand*{\vct}[1]{\boldsymbol{#1}}
\newcommand*{\pd}[2]{\ensuremath{\frac{\partial #1}{\partial #2}}}
\newcommand*{\pdt}[1]{\pd{#1}{t}}
\newcommand*{\dif}{\ensuremath{\mathrm{d}}}
\newcommand*{\Pe}{\ensuremath{\mathrm{Pe}}}
\renewcommand*{\Re}{\ensuremath{\mathrm{Re}}}

\begin{document}

\hypersetup{
  allcolors=black,
  urlcolor=blue,
  linkcolor=blue,
}

\begin{frontmatter}
  \title{\thetitle}
  \date{\today}

  \author[sintef]{Ole H.~H.~Meyer\corref{cor1}}
  \ead{ole.meyer@sintef.no}
  \author[sintef]{Karl Yngve Lervåg}
  \author[sintef]{Åsmund Ervik}
  \address[sintef]{SINTEF Energy Research, P.O. Box 4671 Sluppen, NO-7465 Trondheim, Norway}

  \begin{abstract}
    The numerical simulation of fluid flow through a complex geometry with heat transfer is of strong interest for many applications, such as oil-filled power transformers.
    A fundamental challenge here is that high resolution is necessary to resolve the fluid flow phenomena, but this makes simulation of the full geometry very expensive in terms of computational power.
    In this work, we develop a simulation methodology that combines a porous-medium approach for simulating some regions of the domain, coupled with fully resolved simulations in those regions which are deemed most interesting to study in detail.
    As one does not resolve flow features like thermal boundary layers in the regions modeled with the porous approach, the resolution in these parts can be orders of magnitude coarser.
    This multiscale approach is validated against the use of fully resolved simulations in the whole domain, as well as against analytical solutions to the extended Graetz problem.
    We then apply the approach to the study of oil flow and heat transfer in large electric power transformers and demonstrate a significant reduction in computational cost compared to a fully resolved approach.
  \end{abstract}

\end{frontmatter}

\section{Introduction}
\label{sec:Introduction}
Electric power transformers constitute critical infrastructure, and their safe and steady operation manifests a major task with respect to modern nations' security of energy supply.
While power transformers reach high efficiencies, the power being transformed reaches into the tens or hundreds of mega-volt-amperes (MVA), which is equivalent to megawatts if the load is purely resistive.
This means that even a \SI{1}{\percent} loss dissipated in the transformer represents a large requirement for cooling.
Power transformers in the grid consist of three legs (for three phase power), where each leg is made up of several coaxial windings around a ferrite core, and is several meters tall.
A typical arrangement is to have a low voltage winding, a high voltage winding and a tertiary winding around each core.
Since the power dissipated is proportional to the square of the current, the low voltage winding is usually the main focus for cooling purposes.
The windings are made up of copper that is turned in an overall helical fashion from bottom to top, but the detailed arrangement can be very complex for optimizing the electrotechnical aspects.
The turns are held apart by insulating spacers, and the gaps between turns are filled with oil for the sake of electrical insulation and cooling.
The coaxial windings are also held separated from each other by insulating plates and oil.
An external metal tank contains the three legs and the oil, and supports the total weight of the transformer which is in the tens to hundreds of tons.
On smaller units, the tank may have external fins for cooling of the oil, while on larger units there are dedicated heat exchangers that provide the oil cooling against the ambient air, with or without forced convection on the air side.

\begin{figure}
  \centering
  \begin{subfigure}[b]{0.14\textwidth}
    \centering
    \includegraphics[height=4.5cm]{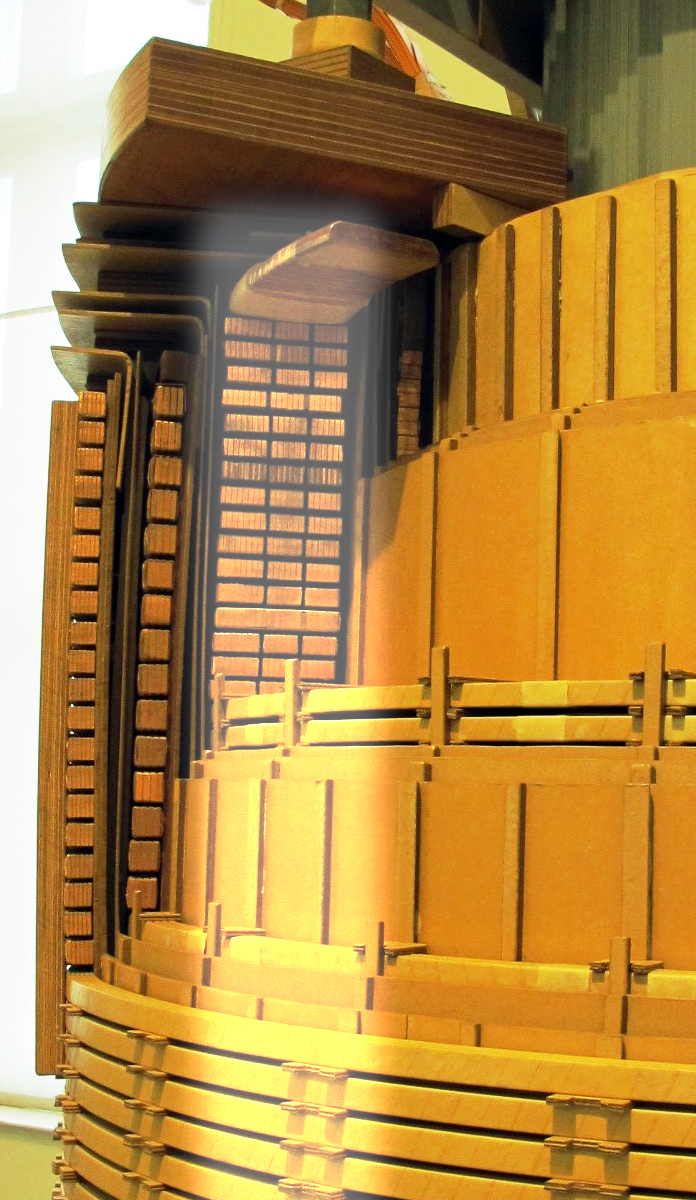}
    \caption{}
    \label{fig:winding-sketch-real}
  \end{subfigure}
  \begin{subfigure}[b]{0.4\textwidth}
    \centering
    \includegraphics[height=4.5cm]{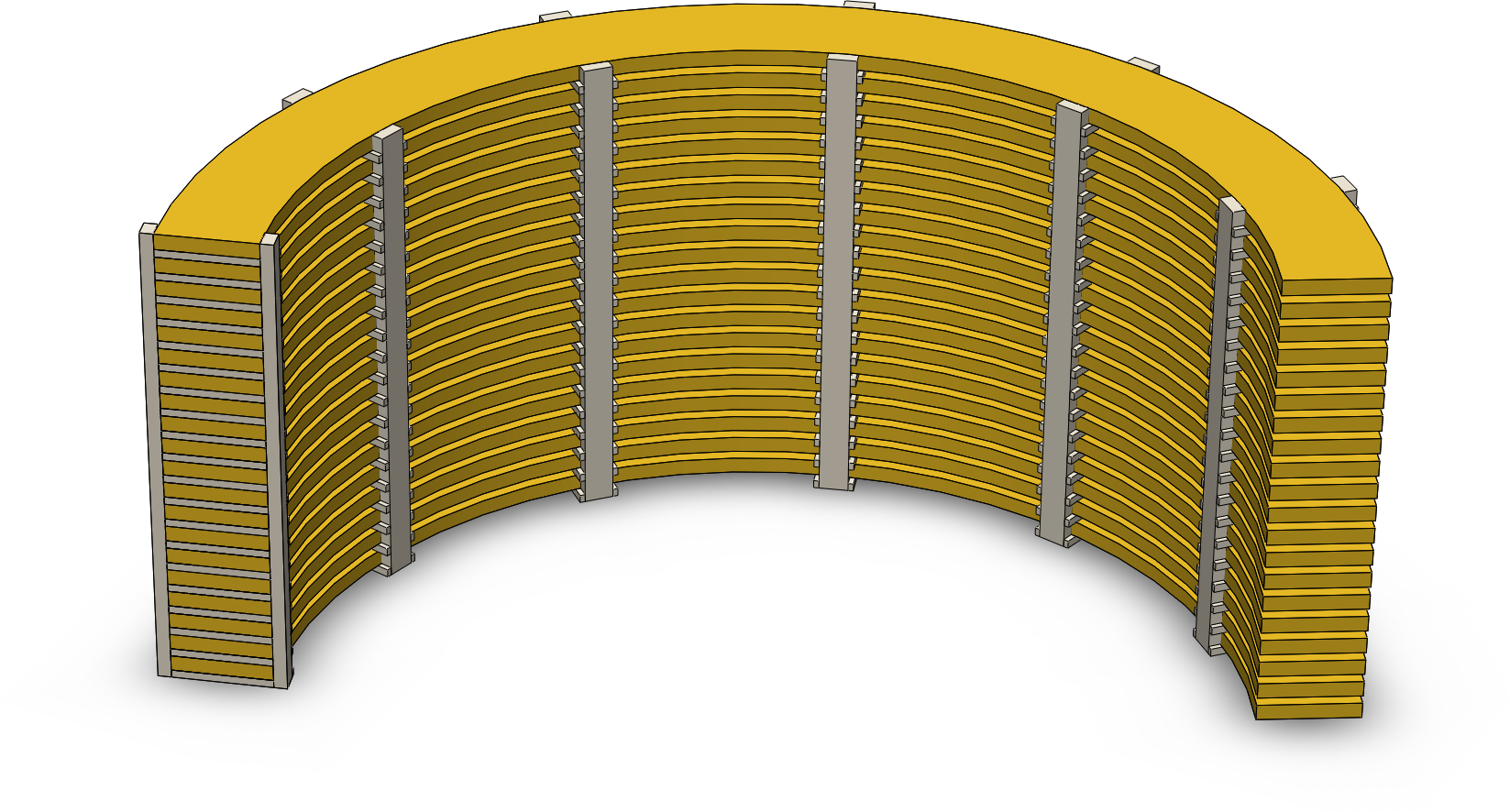}
    \caption{}
    \label{fig:winding-sketch-a}
  \end{subfigure}
  \begin{subfigure}[b]{0.21\textwidth}
    \centering
    \includegraphics[height=5.2cm]{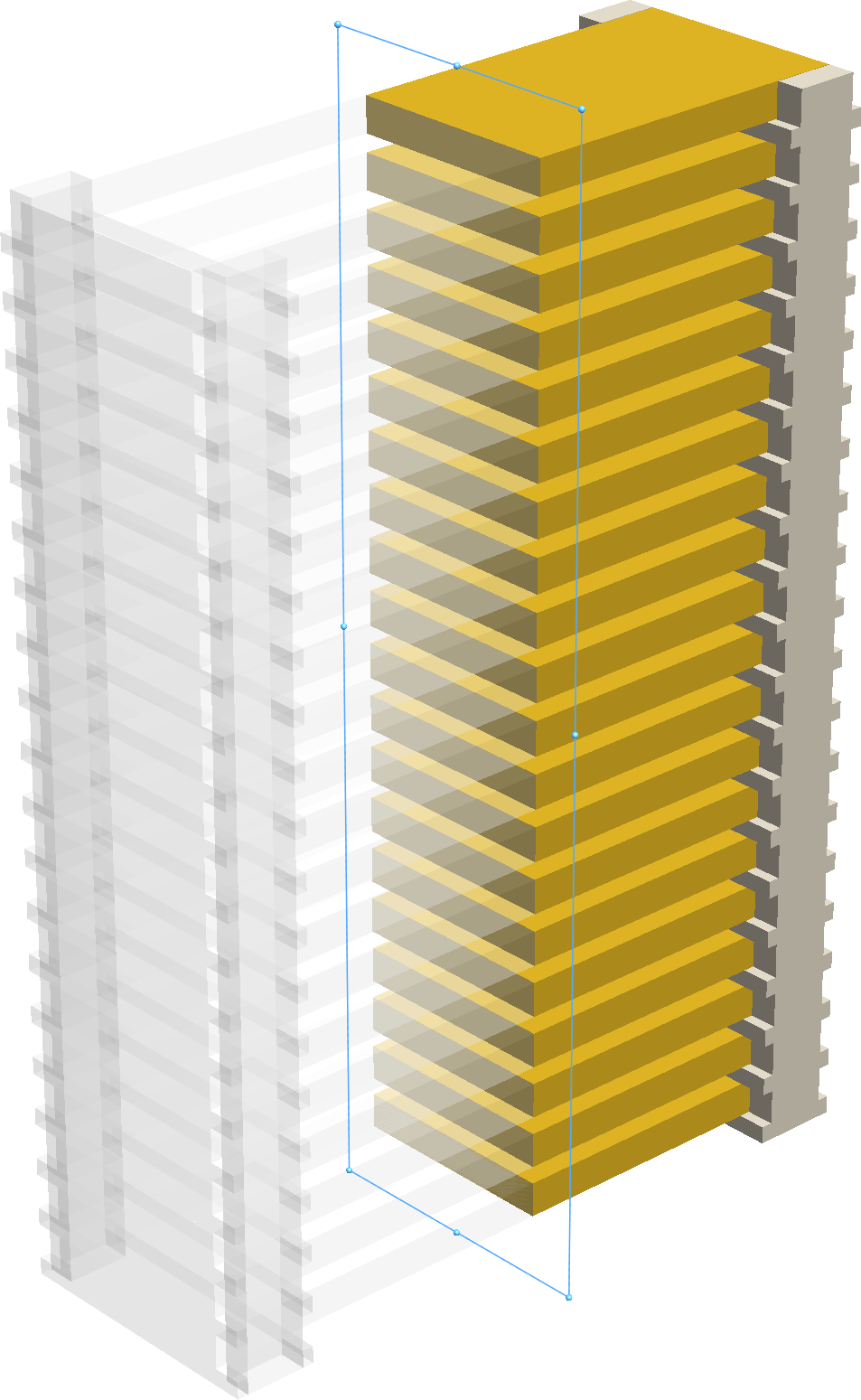}
    \caption{}
    \label{fig:winding-sketch-b}
  \end{subfigure}
  \begin{subfigure}[b]{0.18\textwidth}
    \centering
    \includegraphics[height=5cm]{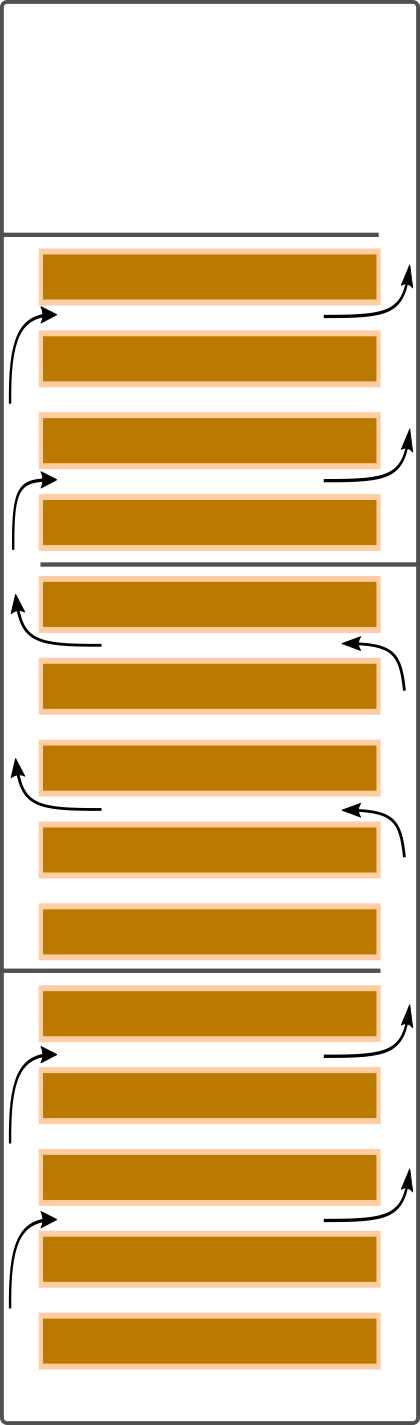}
    \caption{}
    \label{fig:winding-sketch-c}
  \end{subfigure}
  \caption{(a) Cross-section of a smaller 40 MVA transformer, with low voltage winding highlighted.
    (b) Sketch of idealised circular transformer winding.
    (c) Straightened out cross-section between azimuthally adjacent plates.
  (d) Typical zig-zag flow path. [Fig. (a) adapted from a public domain image of a display at Technisches Museum Wien.]}
  \label{fig:winding-sketch}
\end{figure}

\Cref{fig:winding-sketch} shows a section cut of a smaller 40 MVA oil-filled transformer (\cref{fig:winding-sketch-real}), and an idealised transformer winding (half a coil is depicted in \cref{fig:winding-sketch-a}).
To improve cooling by the circulating oil flow, pass washers are installed, resulting in zig-zag motion of the oil as seen in \cref{fig:winding-sketch-c}.
A typical placement of a modelling domain is shown as the blue cross-section in the middle.

The structure of a transformer resembles a heat exchanger, in the sense that it is cooled by allowing the oil to flow between the turns of a winding in a suitable fashion.
However, since the electrical insulation aspects are crucial, it is not feasible to optimize the geometry based on the flow requirements.
Thus the flow patterns are severely restricted, and a typical configuration is that the oil follows a zig-zag path from bottom to the top of the winding.
This means that as one proceeds towards the top of the winding, the incoming oil for cooling has already been heated by previous turns, and thus \emph{hot spots} are often found near the top of the winding.
This is further complicated in that the dissipated heat is not uniform, as in addition to the uniform resistive losses, there are eddy current losses which are typically at their highest near the top.
In addition, the presence of the remaining transformer-legs (dependent on the voltage phasing) affects the azimuthal distribution of power dissipated from the different turns of the windings.

As can be seen from \cref{fig:winding-sketch}, there is very limited space inside the power transformer.
It is therefore difficult to install sufficient instrumentation to find the location and temperature of the hot spot.
Naturally, one would like to perform numerical simulations of the transformer in order to predict the temperature development.
However, the difference in length scales inside the transformer is a major obstacle for such simulations.
The total winding height may be greater than two meters, but the distance between two turns in the winding is on the order of some millimeters.
The smallest distance should be resolved with several tens of grid cells to capture the thermal boundary layers.
Oil properties greatly influence the Prandtl number and consequently the thermal boundary layer width.
Clearly, employing a uniform grid, the resolution requirement would result in astronomical number of necessary grid cells.

Determining both the location and the temperature of hot spots has been the focus of a significant body of research. We do not aim to give an exhaustive review here (see e.g. Ref. \cite{cordoba2019} for a recent overview); previous works focusing on numerical simulations can be found e.g.~in Refs. \cite{kranenborg2008,fdhila2011,laneryd2014,zhang2016,campelo2016,weinlader2012,skillen2012,torriano2012,gastelurrutia_numerical_2011,torriano2018,cordoba2019,santisteban2019,daghrah2020,zhang2020}.
The test codes for power transformers (IEEE C57.12.90, IEC 60076-2), as well as the loading guides for power transformers (IEEE C57.91, IEC 60076-7), are also concerned with the hot spot temperature.
However, previous work has been constrained by the vast computational effort required to consider realistic geometries, so the focus has been either on construction and validation of more simplified thermo-hydraulic network models, or on idealised geometries.

When faced with a problem featuring physical phenomena at scales separated by many orders of magnitude, a powerful technique is to employ a multiscale methodology.
\citet{morega1995} have shown that a stack of heated parallel plates subject to free-stream cooling features an optimal geometry in terms of number of plates and plate spacing.
For suboptimal cooling arrangement (larger number of plates or smaller plate spacing), they demonstrate that overall heat-flux and hot spot temperature can be accurately captured by representing the stack as an anisotropic medium at significantly reduced computational cost.

The application of a porous-media approximation for modelling the thermofluid flow in complex geometries has been widespread for decades and has been summarized in several books, e.g. Refs. \cite{nield1999,lauriat2000,bejan2004}. This includes application to power transformer radiators \cite{paramane2016,kim2018}.
Recently it has been applied also for transformers internals, where \citet{cordoba2019} were the first to apply a porous-medium approach to the three-dimensional simulation of a full power transformer geometry, using their in-house code.
They achieved overall good agreement with experimental results, and showcased that such an approach can give important insights into the global flow features.
A drawback of that approach is that due to some convergence issues, isotropic permeability was used in the winding region, leading to vertical flow velocities inconsistent with experiments.
Earlier work by \citet{gastelurrutia_numerical_2011} considered two-dimensional simulation of the transformer with a porous approximation of some sort, but no details are given concerning the nature or magnitude of the porous resistance, whether it was isotropic or not, or how it was implemented.

In this contribution, we propose a multiscale resolution approach to the problem of thermal flow in transformers.
As hot spots are expected close to the top of the winding, this top region is fully resolved.
The remainder of the geometry is modelled as an anisotropic porous-medium at coarse resolution.
Effectively, the porous region will convey accurate averaged macroscopic flow properties as inflow boundary conditions into the fully resolved domain.
The goal of the approximate model is to provide similar quantities for top-oil and averaged-disk temperature, compared to a fully resolved stack.
This approach promises substantial speedups in terms of computational problem-size reduction, and can enable simulations of the entire three-dimensional transformer geometry with full resolution in the regions of interest.
The main novelty of the present approach is the multiscale combination of a porous approximation in most parts of the domain, which is coupled with highly detailed simulations of the hot-spot region. A further novelty is the detailed derivation of the permeability used in porous simulations, by solving analytically a Poisson equation, which removes the need for time-consuming experiments to characterize the pressure drop as a function of flow rate. We apply a strongly anisotropic permeability, which efficiently cancels out the vertical velocities inside the transformer windings modelled by the porous medium. 

While here we focus on the application of power transformers, we must stress that the presently developed approach is suitable for a broad range of heat exchangers and related devices.
The only requirement necessary for a significant speedup is that focus can be placed on the detailed study of a sub-region of the total domain, and that analytical (or otherwise well-known) expressions for the permeability can be obtained.
In the present work, this sub-region is known \emph{a priori}, but one might well imagine that the fully resolved sub-region is chosen adaptively according to some criterion.

The paper is structured as follows.
In \cref{sec:theory}, we introduce the governing set of equations for thermal fluid flow and the temperature dependent oil properties.
The distinct geometric features of the model are also presented there.
In \cref{sec:method}, we present the proposed method.
In \cref{sec:illustration}, we validate the method against simulations which are fully resolved in all the domain, as well as against analytical solutions to the extended Graetz problem.
In \cref{sec:demonstration}, we demonstrate the application towards modelling of a representative power transformer geometry.
Finally in \cref{sec:conclusion}, some concluding remarks are offered.

\section{Theory}
\label{sec:theory}
A section of multiple heated parallel plates between which fluid can flow, confined by guiding walls on top and bottom with inlet and outlet on alternating sides, is defined to constitute a \emph{pass}.
\cref{fig:winding-sketch-c} depicts three passes in that respect.
In the following, we distinguish three models: \emph{detailed model} (DM), \emph{porous model} (PM), and \emph{porous-approximate model} (PAM).
The DM refers to a conventional, fully resolved, detailed description of a single pass or a stack of passes, depending on context.
The PM refers to a \emph{single} pass porous-medium approximation.
The PM is constructed to approximate the average properties of the single pass DM by modelling the stacked parallel plates as a porous medium.
The PAM refers to a porous-medium approximation of a number of PMs, coupled with a resolved DM pass, i.e. the PAM can be understood as a combination of the DM and the PM.
Below, the relevant evolution equations for mass, momentum and energy are introduced.
The equations are valid for the listed models above, with geometric distinctions achieved through filtering of the corresponding penalization- and source terms in the momentum and energy equation, respectively.

\subsection{Equations}
\label{sub:theory_equations}
In this work, we consider transformer oil as incompressible, with density variations retained via the Boussinesq approximation.
The system at hand is governed by continuity equations for mass, momentum and energy:
\begin{subequations}
  \begin{align}
    \label{eq:mass-conservation}
    \pdt{\rho} + \div (\rho \vct u) &= 0, \\
    \label{eq:momentum-conservation}
    \pdt{\rho \vct u} + \div (\rho \vct u \otimes \vct u \vct)
      &= - \grad p + \rho \vct g
        + \div \left[\mu \left(\grad \vct u
          + (\grad \vct u)^{\mathrm{tr}} \right)\right]
        + \vct f, \\
    \label{eq:energy-conservation}
    \pdt{\rho h} + \div(\rho \vct u h)
      &= \div \left(\frac{k}{c_p} \grad h\right) +  S.
  \end{align}
  \label{eq:governing-eqs}
\end{subequations}
Here $\rho$ is density~(\si{\kilo\gram\per\meter\cubed}), $\vct u$ is velocity~(\si{\meter\per\second}), $h$ is specific enthalpy~(\si{\joule\per\kilogram}), $p$ is pressure~(\si{\pascal}), $\vct g$ is the acceleration of gravity~(\si{\meter\per\second\squared}), $\mu$ is the dynamic viscosity~(\si{\pascal\second}), $k$ is the heat conductivity~(\si{\watt\per\meter\per\kelvin}), and $c_p$ is the specific heat {capacity}~(\si{\joule\per\kilo\gram\per\kelvin}).
Distinction between the DM and PM is given by the penalization-term distribution~(\si{\newton}), $\vct f$, and volumetric heat generation distribution~(\si{\watt\per\meter\cubed}), $S$.
The choice $\vct f = \vct 0$ and $S = 0$ selects the DM, with heat generation set by boundary conditions on the temperature, $T$~(\si{\kelvin}).
In the PM, $\vct f$ and $S$ are constructed as explained in \cref{subsec:porous-approximation}.

We discard contributions of work due to pressure fluctuations in the energy equation, since the fluid is taken as incompressible.
An equation of state is needed to close the above system of equations.
We employ an Oberbeck-Boussinesq density-temperature relation,
\begin{equation}
  \label{eq:EoS}
  \rho = \rho_0 \left[1 - \beta (T - T_0)\right],
\end{equation}
with a reference density, $\rho_0 \equiv \rho(T_0)$ at the reference temperature $T_0$.
$\beta$ is the thermal expansion coefficient of the fluid~(\si{\per\kelvin}).
The fundamental thermodynamic relation between enthalpy, entropy, and pressure is given by
\begin{equation}
  \label{eq:dh}
  \dif h = T \dif s + \dif p / \rho,
\end{equation}
where $s$ denotes specific entropy (\si{\joule\per\kilo\gram\per\kelvin}).
At constant pressure, one obtains from this equation, when combined with the definition of heat capacity and the second law of thermodynamics, the relation
\begin{equation}
  h = \int c_p \dif T,
\end{equation}
where $c_p$ is the specific heat capacity at constant pressure.

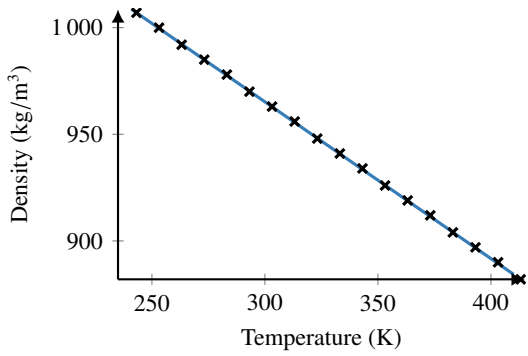
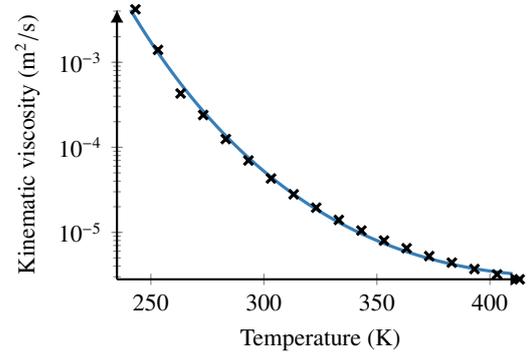
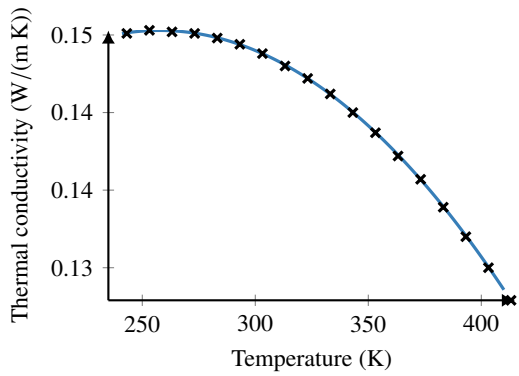
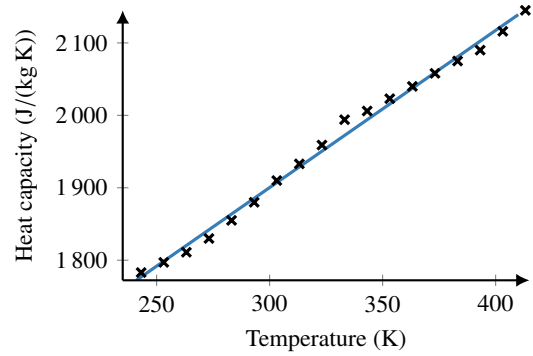
\begin{figure}[tpb]
  \centering
  \pgfplotsset{
    domain=241:410,
    xmin=235,
    xmax=415,
  }
  \begin{subfigure}[b]{0.49\textwidth}
    \centering
    \tikzsetnextfilename{midel-density}
    \begin{tikzpicture}
      \begin{axis}[
          xlabel=Temperature (\si{\kelvin}),
          ylabel=Density (\si{kg/m^3}),
        ]

        \addplot[dp2] table
          [x expr=\thisrow{T}+273.160, y=rho,
          skip first n=4, col sep=semicolon]
          {data/midel.csv};
        \addplot[regression] {1007.0*(1.0 - 7.3e-4*(x - 243.16))};
      \end{axis}
    \end{tikzpicture}
    \caption{Density (\cref{eq:EoS}).}
    \label{fig:rho-verification}
  \end{subfigure}
  \begin{subfigure}[b]{0.49\textwidth}
    \centering
    \tikzsetnextfilename{midel-viscosity}
    \begin{tikzpicture}
      \begin{semilogyaxis}[
          xlabel=Temperature (\si{\kelvin}),
          ylabel=Kinematic viscosity (\si{\meter\squared\per\second}),
        ]

        \addplot[dp2] table
          [x expr=\thisrow{T}+273.16, y expr=\thisrow{nu}/1000000,
          skip first n=4, col sep=semicolon]
          {data/midel.csv};
        \addplot[regression]
          {exp(20.81369191*(ln(x))^2 - 252.81869067*ln(x) + 755.03026555)};
      \end{semilogyaxis}
    \end{tikzpicture}
    \caption{Viscosity (\cref{eq:regr_visc}).}
    \label{fig:viscosity_midel}
  \end{subfigure}

  \vspace{1em}
  \begin{subfigure}[b]{0.49\textwidth}
    \centering
    \tikzsetnextfilename{midel-thermal-conductivity}
    \begin{tikzpicture}
      \begin{axis}[
          xlabel=Temperature (\si{\kelvin}),
          ylabel=Thermal conductivity (\si{\watt\per\meter\per\kelvin}),
        ]

        \addplot[dp2] table
          [x expr=\thisrow{T}+273.16, y expr=\thisrow{k},
          skip first n=4, col sep=semicolon]
          {data/midel.csv};
        \addplot[regression] {-7.2e-7*x^2 + 3.71e-4*x + 9.75e-2};
      \end{axis}
    \end{tikzpicture}
    \caption{Thermal conductivity (\cref{eq:regr_k}).}
    \label{fig:therm_cond_midel}
  \end{subfigure}
  \begin{subfigure}[b]{0.49\textwidth}
    \centering
    \tikzsetnextfilename{midel-heat-capacity}
    \begin{tikzpicture}
      \begin{axis}[
          xlabel=Temperature (\si{\kelvin}),
          ylabel=Heat capacity (\si{\joule\per\kilogram\per\kelvin}),
        ]

        \addplot[dp2] table
          [x expr=\thisrow{T}+273.16, y=c_p,
          skip first n=4, col sep=semicolon]
          {data/midel.csv};
        \addplot[regression] {1249.29 + 2.17*x};
      \end{axis}
    \end{tikzpicture}
    \caption{Heat capacity (\cref{eq:regr_c_p}).}
    \label{fig:heat_cap_midel}
  \end{subfigure}
  \caption{Properties of MIDEL 7131 as a function of temperature.
    The black stars denote measurements from Ref.~\citep{midel2018} and the solid blue lines correspond to the respective regression functions.
  Note that for viscosity, the y-axis is log-scaled.}
  \label{fig:midel}
\end{figure}

In the remainder of the work, we consider fluid data that corresponds to the synthetic ester transformer oil, MIDEL 7131 produced by M\&I Materials.
The physical property data used here are given by the manufacturer \citep{midel2018}.
The thermal expansion coefficient is $\beta = \SI{7.3e-4}{1/\kelvin}$, and we use a reference density $\rho_0 = \SI{1007}{kg/m^3}$ evaluated at $T_0 = \SI{243.16}{\kelvin}$.
This fully specifies the density through the Oberbeck-Boussinesq relation~\eqref{eq:EoS}.
Note that a constant thermal expansion coefficient is sufficient to accurately describe the observed linear relation among density and temperature, cf. \cref{fig:rho-verification}.
The remaining properties are specified through the following regression functions,
\begin{align}
  \label{eq:regr_visc}
  \ln \nu(T) &= 20.81369191\ln^2 T - 252.81869067\ln T + 755.03026555, \\
  \label{eq:regr_k}
  k(T) &= -\num{7.2e-7} T^2 + \num{3.71e-4} T + \num{9.75e-2}, \\
  \label{eq:regr_c_p}
  c_p(T) &= 2.17T + 1249.29,
\end{align}
where $\nu = \mu / \rho$ is the kinematic viscosity~(\si{\meter\squared\per\second}).
See \cref{fig:midel} for comparisons with measurements.

\subsection{Geometric dimensions}
\label{subsec:dimensions}
\begin{figure}[bt]
  \centering
  \includegraphics{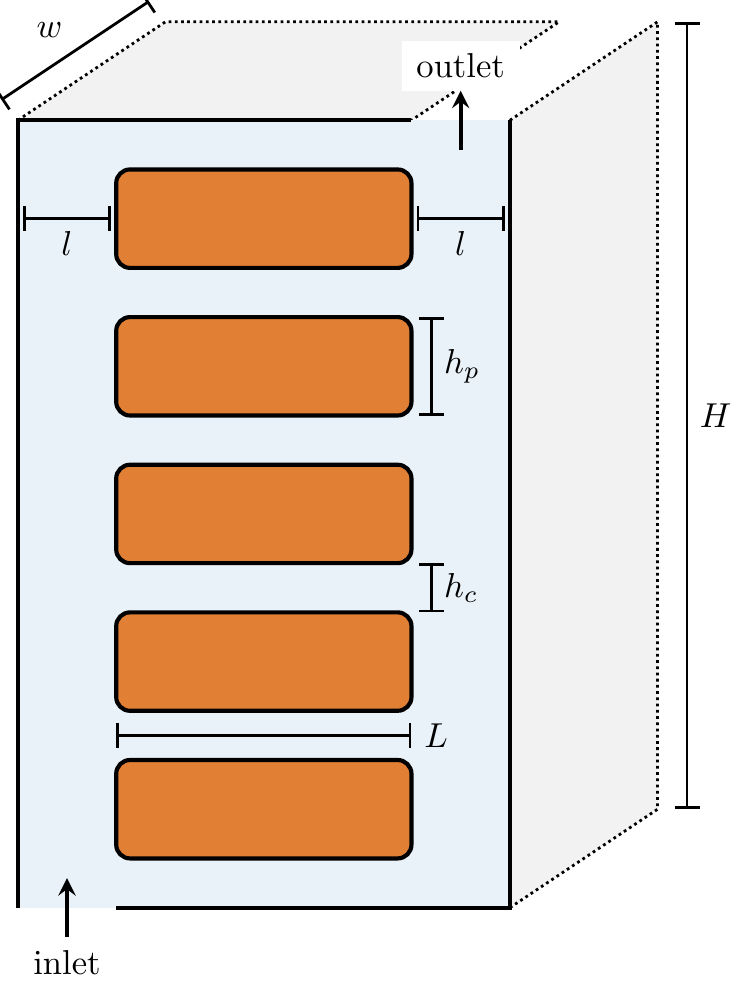}
  \caption{A sketch of the generic computational domain of a single, detailed pass with the characteristic dimensions and inlet/outlet denoted.
    The number of plates, $N$, in a pass is typically larger than depicted.
    Here $H$ is the pass height, $L$ is the channel length, $h_p$ is plate height, $h_c$ is channel height, and $l$ is the left and right leg lengths.
  The pass has a depth $w$ into the paper plane.}
  \label{fig:geometry}
\end{figure}

The governing equations are solved in different geometric domains for the DM, PM, and PAM, respectively.
The DM is solved on one or more detailed passes, where a detailed pass is a ``high-resolution zoom'' that resolves the full transformer pass geometry, see \cref{fig:geometry}.
A detailed pass consists of $N+1$ channels between the turns, where each channel has length $L$ and height $h_c$.
The distance between the channels is $h_p$ and the thickness of the left and right legs is $l$.
$w$ denotes the depth of the pass.
For the present purpose, we assume equidistant channel heights for all $N+1$ channels in the pass.
Next, the PM is solved on one or more porous blocks.
A porous block is a cuboid with dimensions $L_p\times H\times w$, where $L_p$ and $H$ is related to the detailed pass through $L_p = 2l + L$ and $H = N h_p + (N+1) h_c$.
Finally, the PAM is solved on three porous blocks stacked atop each other, with a single detailed pass on top.

In \cref{subsec:analytical-comparison,subsec:turning-flow}, we consider geometries that can be classified as suboptimal in the spirit of \citet{morega1995}:
Both geometries feature more plates and reduced plate distance, compared to a configuration that would maximize the free stream cooling.
Consequently, the cases at hand constitute good candidates for approximation by an anisotropic porous medium.
Though fluid properties and flow pattern in the present contribution are different from the idealized setting in Ref.~\citep{morega1995}, we demonstrate in the following sections that the porous-medium approximation indeed is a viable approach for assessing thermal flow properties in a typical transformer winding geometry.

\subsection{Porous media approximation}
\label{subsec:porous-approximation}
The PM is constructed by placing a porous medium of dimensions $L\times w \times H$ into the porous block of dimensions $L_p \times w \times H$.
The (remaining) left and right legs serve as transition domains, i.e.~the distributions $\vct f$ and $S$ vanish in the legs.
Modelling of the underlying porous medium is contained as velocity-proportional drag in the penalization term~$\vct f$ in \cref{eq:momentum-conservation}:
\begin{equation}
  \label{eq:source}
  f_{i} = - \mu D_{i j} u_j - \frac{1}{2} |u_{k k}| F_{i j} u_j,
\end{equation}
where $D_{i j}$ and $F_{i j}$ are the Darcy and Forchheimer coefficients, respectively.
The latter is relevant for high Reynolds number flows and neglected in the following \citep{forchheimer1901}.

The Darcy coefficient is given by
\begin{equation}
  D_{i j} = \kappa_i^{-1} \delta_{i j},
\end{equation}
with the $i$-component of the permeability tensor, $\kappa_i$~(\si{\meter\squared}), and the Kronecker delta, $\delta_{i j}$, which equals unity for like indices and vanishes for cross-directions.
One can derive Darcy's law from the Navier-Stokes momentum balance (\cref{eq:momentum-conservation}) with the penalization term in \cref{eq:source} if one assumes that the drag force is large compared to the inertia term and the divergence of the stress tensor, cf.~\citep{narsilio2009}.
We provide details on the calculation of the permeability tensor in \cref{subsec:permeability}.

With the momentum penalization in place, it remains to devise a proper heat-source distribution.
In this work, we only consider the fluid part of the problem, thus the temperature evolution in the porous block is given as~\citep{nield1999}
\begin{equation}
  \label{eq:porousTemp}
  \frac{\partial T}{\partial t}
  + \nabla \cdot \left( \frac{1}{\phi}\vct{u} T \right)
  = \nabla \cdot \left( \mathbf{a} \nabla T \right) + s',
\end{equation}
where $\phi$ denotes the porosity~(-) and $s'$ (\si{\kelvin\per\second}) denotes a generic volumetric heat source.
In this context, the porosity is given by the ratio of the fluid domain and the total volumes.
The heat diffusivity tensor $\mathbf{a}$~(\si{\meter\squared\per\second}) features distinct components along its diagonal to account for the structure of the detailed pass.
Its off-diagonal elements are zero.
From the analogy between electric resistivity and thermal resistance, we find that heat conduction parallel to the channels (in the $x$ and $y$ directions), is $a_x = a_y = \phi \alpha$, where $\alpha = k / (\rho c_p)$ denotes thermal diffusivity (\si{\meter\squared\per\second}).
Similarly, heat conduction across the channels yields $a_z = \alpha / \phi$.
Volumetric heat production is contained in the source term $s'$, which should be understood in terms of the heat generation rate from the coils.
\cref{eq:porousTemp} can be rewritten as an enthalpy balance equation
\begin{equation}
  \label{eq:porousTempEnth}
  \frac{\partial \rho h}{\partial t}
  + \nabla \cdot \left( \frac{\rho}{\phi}\vct{u} h \right)
  = \nabla \cdot \left( \frac{\mathbf{k}}{c_p} \nabla h \right) + S,
\end{equation}
with the heat-conductivity tensor $\mathbf{k}$~(\si{\joule\per\kilo\gram\per\kelvin}), and the source term $S$ now given in \si{\watt\per\meter\cubed}.
In steady state, the total heat provided by the source term is given by
\begin{equation}
  \phi S V_\mathrm{por} =
      \int \rho \vct u h \cdot \mathrm{d} \vct A_\mathrm{inlet}
    - \int \rho \vct u h \cdot \mathrm{d} \vct A_\mathrm{outlet},
\end{equation}
with $V_\mathrm{por} = L w H$ denoting the volume of the porous region where the source term is active.
Similarly, integrating \cref{eq:energy-conservation}, and requiring the heat fluxes into and out of the PM and the DM to equate, we find
\begin{equation}
  \label{eq:source-term}
  S = \frac{k(\grad T)\bigr\rvert_\text{coil}
    \:A_\text{coil}}{\phi V_\text{por}}.
\end{equation}
Here, the heat-flux at the coil boundaries is controlled by a constant temperature gradient, $(\grad T)|_\text{coil}$, which is active on the coil area $A_\text{coil}$.
We note that the enthalpy advection velocity in the porous formulation is at the Darcy velocity, $\vct u / \phi$.
Details on how the equations are solved simultaneously on the resolved and porous domains are provided in \cref{sec:method}.

\subsection{Permeability for stacked parallel planes}
In the following, we derive expressions for the permeability.
At its core, the derivations rely on solving the steady-state velocity profile in the laminar and incompressible limit of the Navier-Stokes equations.
This corresponds to solution of a Poisson problem,
\begin{equation}
  \label{eq:steady-state-velocity}
  \grad p = \mu \grad^2 \vct u,
\end{equation}
subject to appropriate boundary conditions.
\Cref{eq:steady-state-velocity} is also referred to as the Hagen--Poiseuille flow problem.
Integration of the velocity over an area perpendicular to the flow direction, gives the associated volumetric flow rate, $Q$ (in \si{\meter\cubed\per\second}),
\begin{equation}
  Q = \int \vct u \cdot \mathrm{d} \vct A. 
\end{equation}
\label{subsec:permeability}
The volumetric flow rate for a single channel $i$ of height $h_i$ is, according to parallel plane Pouiseuille flow,
\begin{equation}
  \label{eq:pouiseuille}
  Q_i = - \frac{h_i^3 w}{12 \nu} \frac{\Delta p}{L},
\end{equation}
where $\Delta p$ is the density weighted pressure difference along channel $i$, i.e. $p \rightarrow p / \rho$.
Note that \cref{eq:pouiseuille} follows from the no-slip boundary condition on the velocity in the vertical direction (at the plates) and slip boundary condition in the horizontal direction (symmetry).
According to Darcy's law, the volumetric flow rate due to a pressure gradient is:
\begin{equation}
  \label{eq:DarcyFlowRate}
  Q_\mathrm{D} = - \frac{\kappa A}{\nu} \frac{\Delta p}{L} = -\frac{\kappa w H}{\nu} \frac{\Delta p}{L},
\end{equation}
where $A = w H$ is the area perpendicular to the flow direction.
Consequently, we find the effective permeability for $N$ channels of aperture $h_i = h_c \forall i$ upon equating $ \sum Q_i = Q_\mathrm{D}$:
\begin{equation}
  \kappa = \frac{N h_c}{H}\frac{h_c^2}{12} = \phi \kappa',
  \label{eq:slipPermeability}
\end{equation}
in terms of the porosity $\phi = N h_c / H$ and single-channel permeability $\kappa' = h_c^2 / 12$.

The above relation for the volumetric flow rate~\eqref{eq:pouiseuille} is valid for $h_c/w \ll 1$ or symmetry of the flow field beyond the channel width.
Finite channel widths with no-slip boundary conditions on the velocity in both transversal directions can be resolved by performing a Fourier series expansion of the velocity field in the Hagen--Poiseuille flow problem (cf. \cref{eq:steady-state-velocity} ).
It follows that the volumetric flow rate for a rectangular channel of finite aspect ratio is given by
\begin{equation}
  \label{eq:FiniteAspectVolumeRate}
  Q_i = \frac{w h_i^3 \Delta p}{12 \nu L} \left[1 - \sum_{n, \mathrm{odd}} \frac{192}{\pi^5} \left( \frac{h_i}{w} \right) \frac{1}{n^5} \tanh \left( \frac{n \pi w}{2 h_i} \right) \right],
\end{equation}
where the index label ``$n, \mathrm{odd}$'' denotes odd integers, $n \in \{1, 3, 5, ...\}$.
Accordingly, the effective permeability for $N$ channels of finite width is given by
\begin{equation}
  \label{eq:FiniteAspectPermeability}
  \kappa = \frac{N h_c}{H} \frac{h_c^2}{12} \left[1 - \sum_{n, \mathrm{odd}} \frac{192}{\pi^5} \left( \frac{h_c}{w} \right) \frac{1}{n^5} \tanh \left(\frac{n \pi w}{2 h_c} \right) \right].
\end{equation}

The impact of the no-slip boundary conditions in the transverse direction on the steady-state flow rate is vanishingly small as $h_c/w \ll 1$, where \cref{eq:FiniteAspectPermeability} approaches \cref{eq:slipPermeability}.
Typically, this condition is well satisfied.
For instance, Ref.~\citep{cigre2016} discusses a transformer leg of radius $\sim \SI{300}{\milli\meter}$ with channel height $\SI{4}{\milli\meter}$, divided in 18 azimuthal sections.
In that case, $h_c / w \sim 4/100$, and \cref{eq:slipPermeability} is recovered.
It follows that we expect pure 2D simulations to constitute a decent approximation to the flattened 3D setting.
We employ \cref{eq:slipPermeability} for 2D simulation.

\subsection{Permeability of stacked annulus segments}
The full 3D case retains curvature.
This in turn mandates to solve the Poisson equation, \cref{eq:steady-state-velocity}, for the steady-state volumetric flow rate on an annular domain.
In cylindrical coordinates, we take the annular channel to extend radially $\Delta R = R_2 - R_1$, axially $h = z_2 - z_1$, and azimuthally $\Delta \theta = \theta_2 - \theta_1$.

It can be shown that the radial velocity field is given by
\begin{equation}
  \label{eq:radial-flow}
  u_r = \frac{1}{r} \frac{16 \Delta P}{\mu} \sum_{m, \mathrm{odd}} \sum_{n, \mathrm{odd}} \left[\frac{m^2}{2 (\Delta \theta)^2}\left(\frac{1}{R_1^2} - \frac{1}{R_2^2}\right) + \frac{n^2}{h^2} \log \left(\frac{R_2}{R_1} \right) \right]^{-1} \sin \left(\frac{m \pi \theta}{\Delta \theta} \right) \sin \left(\frac{n \pi z}{h} \right),
\end{equation}
with the radial pressure drop $\Delta P = P_1 - P_2$, and $P_1 = p (R_1), P_2 = p (R_2)$.
The permeability is related to the radial flow rate via Darcy's law
\begin{equation}
  \label{eq:Darcy-radial}
  Q_r = \frac{\kappa \Delta \theta h}{\mu} \frac{\Delta P}{\log \left(R_2 / R_1 \right)}.
\end{equation}
Integrating the radial velocity from \cref{eq:radial-flow} over azimuthal and axial directions provides the radial flow rate, $Q_r = \int r u_r \mathrm{d}\theta \mathrm{d}z$, which equated with the radial Darcy law, \cref{eq:Darcy-radial}, yields an expression for the permeability characteristic of $N$ vertically stacked annular segments of height $H$:
\begin{equation}
  \label{eq:radial-permeability}
  \kappa = \frac{64 N}{\pi^2} \left(\frac{h}{H} \right) \log \left(\frac{R_2}{R_1} \right) \sum_{m, \mathrm{odd}} \sum_{n, \mathrm{odd}} \left[\frac{m^2}{2 (\Delta \theta)^2}\left(\frac{1}{R_1^2} - \frac{1}{R_2^2}\right) + \frac{n^2}{h^2} \log \left(\frac{R_2}{R_1} \right) \right]^{-1} \frac{1}{m n}.
\end{equation}

\section{Numerical methods and implementation}
\label{sec:method}
We use the open-source framework OpenFOAM to solve the governing equations.
OpenFOAM provides a generic framework for finite-volume discretization of partial differential equations.
It is written as a set of C++ libraries, and its object-oriented structure allows for close top-level representation of the mathematical formulations.
This enables intuitive custom development and modification~\citep{weller1998}.
The flexibility of OpenFOAM for tailor-made applications has received increasing attention recently \citep{jasak2009}.

A typical workflow consists of specifying initial and boundary conditions for the field variables at hand in separate files, as well as mesh files that contain the discretization domain and configuration files to specify the solver with numerical schemes and solution/convergence criteria.
This work employs the PIMPLE algorithm for pressure-velocity coupling.
The PIMPLE algorithm is a hybrid SIMPLE--PISO iteration scheme that allows larger time steps.
Adaptive time steps limited by a user defined Courant number\footnote{The Courant number gives a necessary stability condition that relates the time step and spatial discretization. It can be understood as a constraint on the minimum allowable propagation velocity of numerical waves. We also refer to the Courant number as the CFL number.} may be chosen.
Summarized, the SIMPLE algorithm~\citep{patankar1972} contains the following steps:
\begin{enumerate}
  \item solve for the velocity vector from the momentum equation with an initial guess of the pressure
  \item add corrections to the velocity and pressure
  \item solve for the pressure corrections
  \item solve for the velocity corrections
  \item repeat until the convergence criterion is reached
\end{enumerate}
The PISO algorithm~\citep{issa1985} adds a second corrector stage to obtain better convergence.
Time integration is performed by an implicit, first-order Euler scheme.
Interpolation of the face fluxes to the cell values is achieved by combinations of second-order central-differencing schemes.
At each time step, the convergence of velocity, pressure, and enthalpy is monitored.
The algorithm is deemed converged upon reaching residuals of \num{1e-5}.

We have constructed a transient solver \verb+transformerFoam+ based on the standard, transient compressible solver \verb+buoyantPimpleFoam+, with consistent handling of thermophysical properties.
The presence of a porous medium can be represented by modifying the momentum equation of the solver via specification of the tensors $\vct{D}$ and $\vct{F}$ via the \verb+fvOptions+ functionality, which applies the penalization on a defined cell region of the mesh.
Additionally, heat generation is added as a volumetric source term to the energy equation via the \verb+fvOptions+ environment.

The geometry shown in \cref{fig:geometry} is constructed with a standard \verb+blockMeshDict.m4+ file.
A \verb+cellZone+ \emph{porosity} has been defined for the central (porous) part of the PM block.
The velocity and pressure calculations in the porous blocks are handled by adding a penalization term in the \verb+fvOptions+ for the momentum equation over the \emph{porosity} zone of the mesh.
Still, care must be taken for correct application of the temperature equations.
As \cref{eq:energy-conservation} and \cref{eq:porousTempEnth} differ by more than a source term (i.e. same structure with different coefficients), the approach for the momentum equation cannot be applied.
We have chosen to implement the distributions $\epsilon$ and $\delta = 1 - \epsilon$, where $\epsilon$ is equal to one in \emph{porosity} and zero elsewhere.
The enthalpy equations are then solved together in the form
\begin{equation}
  \pdt{ \rho h} + \left(\delta + \epsilon \frac{1}{\phi} \right) \div (\rho \vct{u} h ) = \delta \div \left(\frac{k}{c_p} \grad h \right) + \epsilon \left[ S + \div \left( \frac{\mathbf{k}}{c_p} \grad h \right) \right].
\label{eq:distrTemp}
\end{equation}
Observe that energy is conserved separately in each domain:
In the porous domain where $\delta \rightarrow 0$, \cref{eq:distrTemp} reduces to \cref{eq:porousTempEnth}, and in the resolved domain where $\epsilon \rightarrow 0$, \cref{eq:distrTemp} reduces to \cref{eq:energy-conservation} without the source term.

\Cref{eq:energy-conservation} describes isotropic heat conduction, with the scalar heat diffusivity $\alpha$ occurring in the conduction term.
As the porous medium to be considered is highly asymmetrical, the thermal diffusivity $\mathbf{a}$ in \cref{eq:porousTemp} needs to be cast in tensorial form.
Accordingly, the solver is supplied a tensorial heat diffusivity of type \verb+dimensionedTensor+.
The porous zone is specified by adding \verb+explicitPorositySource+ to the \verb+fvOptions+ file.
This adds a term that correspond to \cref{eq:source} to the momentum equation.
Similarly, \verb+scalarSemiImplicitSource+ represents the heat source and corresponds to \cref{eq:source-term}, applied to the temperature equation.

Temperature-dependent oil properties  are accounted for by constructing a custom \verb+thermophysicalModel+ with fluid properties given in \cref{eq:regr_visc}, \cref{eq:regr_k} and \cref{eq:regr_c_p}.

\section{Illustration of the approach}
\label{sec:illustration}
In this section, we illustrate the core features of the chosen approach.
The task at hand is to \emph{construct} a porous block such as to reproduce the overall flow state of the detailed geometry.
When using numerous porous blocks stacked atop each other, we demand that the throughput of i) the pressure differential from inlet to outlet, ii) the mass flux, and iii) the heat flux align with the detailed geometry.
The constraint of similar mass- and heat fluxes inherently provides similar average temperatures at the outlet of the domain.

We start by validating numerical simulations of detailed and porous blocks in two dimensions.
Physically, this situation resembles a cross cut far from confining lateral boundaries (block washers), such as illustrated by \cref{fig:winding-sketch-b}.

\subsection{Steady-state profiles}
\label{subsec:steady-state-profiles}
\Cref{eq:pouiseuille} and \cref{eq:DarcyFlowRate} can be used to quantify the steady-state pressure-velocity coupling of the numerical calculation.
It is desirable to obtain quantifiable relations for the temperature in both the detailed and porous passes.
We first consider 2D normal inflow into the left boundary in a generic geometry as sketched in \cref{fig:geometry} for the DM.
$10$ channels of height $h_c = \SI{4}{\milli\meter}$ and length $L =  \SI{75}{\milli\meter}$ are employed.
The PM consists of a cuboid of corresponding dimensions as explained in \cref{subsec:dimensions}.
Thermophysical properties are assumed to be temperature independent in this subsection.
In the DM, a parabolic profile for the transverse velocity component inside the channels is set up ($x$ and $z$ are the downstream and transverse coordinates, respectively): $u(z)=4u_\mathrm{m}z(h-z)/h^2$, with the maximum velocity on the centerline given by $u_\mathrm{m} = \Delta P h^2 / 8 \nu L$.
Subject to slip-velocity boundary conditions at the top and bottom wall, the corresponding velocity profile in the porous block is given by the plug flow velocity $u_p = u_0 / \phi$, in terms of the inlet velocity $u_0$ and the porosity $\phi$.

Neglecting axial conduction, the steady state temperature profile satisfies the nonhomogeneous boundary value-problem
\begin{subequations}
  \begin{align}
    \partial_x T &= \gamma \partial_z^2 T + s', \\
    \partial_z T (x,0) = \partial_z T (x,\mathcal{H}) &= f, \\
    T(0,z) &= T^\mathrm{in},
  \end{align}
\end{subequations}
where $\gamma = a_z / u$ with the $z$-dependent velocity given by the parabolic profile in the detailed case, and the uniform plug-flow velocity in the porous case.
For the detailed case, the source term $s' = 0$ and heat generation is due to heat provided from the coils at $z = \mathcal{H} = h_c$, with $f = \partial_n T|_\text{boundary} = (\grad T)|_\text{coil}$.
Here $\partial_n T|_\text{boundary}$ denotes the temperature gradient normal to the boundary, evaluated at the boundary. 
For the porous approximation, there is no influx at the boundaries at $z = \mathcal{H} = H$, leaving $f = 0$, but volumetric heat generation, $s' = S / \rho_0 c_p u_0$, with $S$ according to \cref{eq:source-term}.
This is equivalent to a homogeneous (detailed) or non-homogeneous (porous) diffusion problem in $z$ direction ($x \rightarrow t$) with non-homogeneous (detailed) or homogeneous (porous) boundary conditions and initial value $T^\mathrm{in}$.

In case of two-dimensional parabolic channel flow (i.e. hydrodynamically fully developed) with uniformly applied heat-flux at the transverse boundaries, one encounters an extended Graetz problem, whose solution for the thermally developing channel wall temperature is given by\footnote{This is also valid for wide channels, $h_c/w \rightarrow 0$, with no-slip boundary condition in the depth-direction.}
\begin{equation}
  T^\mathrm{wall} = T^\mathrm{in}
    + \frac{q_\mathrm{wall} a}{k_\mathrm{fluid}} \left[
      \frac{4}{\Pe} \frac{x}{a}
      + \frac{17}{35}
      + \sum_{n=1}^\infty c_n Y_n(1)
        \exp\left(-\frac{8}{3} \frac{\beta_n^2}{\Pe} \frac{x}{a}\right)
  \right].
  \label{eq:detailedTprofile}
\end{equation}
Here, $x$ measures the downstream distance inside the parallel-plate channel and $a = h_c/2$ is the channel half-width, whence $q_\mathrm{wall} = k (\grad T)|_\mathrm{coil}$ denotes the wall heat-flux.
The first three eigenvalues $\beta_n$ and associated values for $Y_n(1)$, as well as asymptotic relations for large $n$ can be found in \citep{cess1959}.
Eigenvalues and values for the eigenfunctions evaluated at the wall, up to $n=10$ can be found in \citep{sparrow1963}.

The porous case is far simpler, as the velocity field is uniform in $z$-direction:
\begin{equation}
  T(x,z) = T^\mathrm{in} + s' x.
  \label{eq:porousTprofile}
\end{equation}
It should be noted that, for this particular boundary value-problem, the eigenvalues and eigenfunctions are given by $\lambda_n = (n \pi / H)^2$ and $\phi_n (z) = \cos \sqrt{\lambda_n} z)$ for $n = 0,1,2, \dots$, annihilating the conventional exponential terms arising in solutions of the heat equation\footnote{Integrals over the eigenfunctions over the width of the channel vanish: $\int_0^H \cos \left(n \pi z / H \right) \mathrm{d}z = 0$}.

\subsection{Comparison to analytical predictions}
\label{subsec:analytical-comparison}
In the following, we consider an idealized detailed geometry with constant and uniform inlet flow from the left and uniform heat flux trough the channel walls.
The front and back channel walls are taken adiabatic for comparison with the Graetz problem.
We consider a pass of $N = 10$ plates with dimensions $L = \SI{75}{\milli\meter}$, $l = \SI{12.5}{\milli\meter}$, $h_c = \SI{2}{\milli\meter}$, $h_p = \SI{8}{\milli\meter}$, and $w = \SI{10}{\milli\meter}$.
The width of the top and bottom channels are half the width of the remaining channels ($h_c$).
We also consider a corresponding porous block.

The thermophysical state of the fluid is assumed isothermal and the respective oil properties are:
Kinematic viscosity $\nu = \SI{4.27e-5}{\meter\squared\per\second}$, density $\rho = \SI{9.6e2}{\kilogram\per\meter\cubed}$, thermal conductivity $k = \SI{1.5e-1}{\watt\per\meter\per\kelvin}$, specific heat capacity at constant pressure $c_p = \SI{1.9e3}{\joule\per\kilogram\per\kelvin}$.
A typical mass flow rate at the inlet, $\dot{m} = \SI{2.167e-3}{\kilogram\per\second}$, is considered.
This is the same as one of the cases considered in \citep{skillen2012}, and a representative value for oil flow in insulated transformers \citep{cigre2016}.
Taking the channel length $L$ as the characteristic spatial scale, the thermofluid is characterized by Reynolds number $\Re = u_\mathrm{mean} L / \nu = 4$, Prandtl number $\Pr = \nu / \alpha = 520$, and Péclet number $\Pe = \Re \Pr = 2080$.
The thermal diffusivity is $\alpha = k / (\rho c_p) = \SI{8.2e-8}{\meter\squared\per\second}$.
Note that the Reynolds and Péclet numbers are evaluated at the mean channel velocity, which for the porous case is given by $u_\mathrm{mean} = u_\mathrm{inlet} / \phi$.
The large Péclet number indicates that heat conduction is essentially negligible compared to heat advection.
For the channels under consideration, we find that the flow field is hydrodynamically developed (the hydrodynamic entrance length $x_\mathrm{en}^\mathrm{hy} = 0.05 D_H \Re_{D_H} \approx \num{2e-3} L$, with the Reynolds number evaluated at the hydraulic diameter) and thermally developing (the thermal entrance length $x_\mathrm{en}^\mathrm{th} = 0.05 \Re_{D_H} \Pr \approx 306 L$).
Gravity is neglected and uniform velocity loads onto the channels are obtained by invoking slip boundary conditions on the top and bottom walls.
The porous block also features slip velocity constraints on the top and bottom boundary.

This case is analogous to the extended Graetz problem discussed in \cref{subsec:steady-state-profiles}, to which an analytical solution for the downstream wall temperature profile exists.
The detailed simulations are verified against this analytical solution to assess necessary grid cells inside the channel domain.
We then construct the corresponding porous block and show that the required constraints on its output can be satisfied at reduced total grid count.

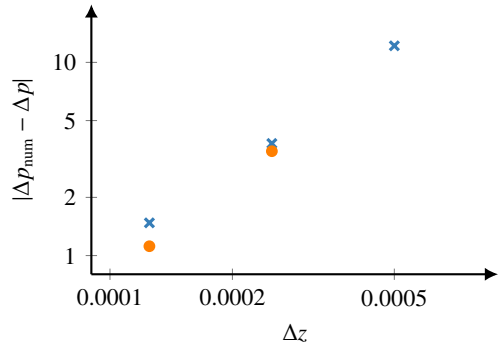
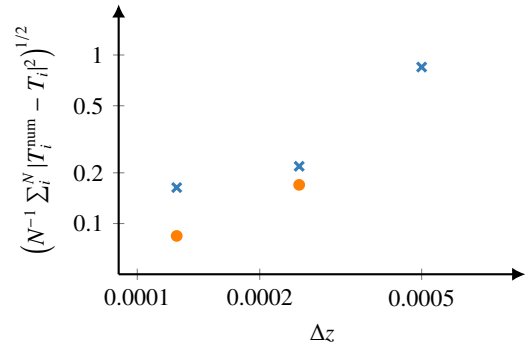
\begin{figure}[tpb]
  \centering
  \begin{subfigure}[b]{0.49\textwidth}
    \centering
    \tikzsetnextfilename{graetz-error-pressure}
    \begin{tikzpicture}
      \begin{loglogaxis}[
          xlabel = $\Delta z$,
          ylabel = $|\Delta p_\text{num} - \Delta p|$,
          xmin = 0.9e-4,
          xmax = 9.0e-4,
          xtick = {1e-4, 2e-4, 5e-4},
          ymin = 0.8,
          ymax = 20,
          ytick = {1, 2, 5, 10},
          log ticks with fixed point,
          table/x = dz,
        ]

        \addplot[dp2, c21] table [y=l1_coarse] {data/graetz-errors.csv};
        \addplot[dp1, c22] table [y=l1_fine]   {data/graetz-errors.csv};
      \end{loglogaxis}
    \end{tikzpicture}
    \caption{L1 error of pressure drop across micro-channels.}
    \label{fig:graetz-error-norms-1}
  \end{subfigure}
  \begin{subfigure}[b]{0.49\textwidth}
    \centering
    \tikzsetnextfilename{graetz-error-temperature}
    \begin{tikzpicture}
      \begin{loglogaxis}[
          xlabel=$\Delta z$,
          ylabel=$\left(N^{-1}\sum_i^N |T_i^\text{num} - T_i|^2\right)^{1/2}$,
          xmin = 0.9e-4,
          xmax = 9.0e-4,
          xtick = {1e-4, 2e-4, 5e-4},
          ymin = 0.05,
          ymax = 2,
          ytick = {0.1, 0.2, 0.5, 1.0},
          log ticks with fixed point,
          table/x = dz,
        ]

        \addplot[dp2, c21] table [y=l2_coarse] {data/graetz-errors.csv};
        \addplot[dp1, c22] table [y=l2_fine]   {data/graetz-errors.csv};
      \end{loglogaxis}
    \end{tikzpicture}
    \caption{L2 error of the channel wall temperature.}
    \label{fig:graetz-error-norms-2}
  \end{subfigure}
  \caption{Error estimates for (a) pressure drop, and (b) channel wall temperature.
  Results are presented for two different downstream grid resolutions: $\Delta x = \SI{5e-1}{\milli\meter}$ (blue stars) and $\Delta x = \SI{2.5e-1}{\milli\meter}$ (orange circles).}
  \label{fig:graetz-error-norms}
\end{figure}

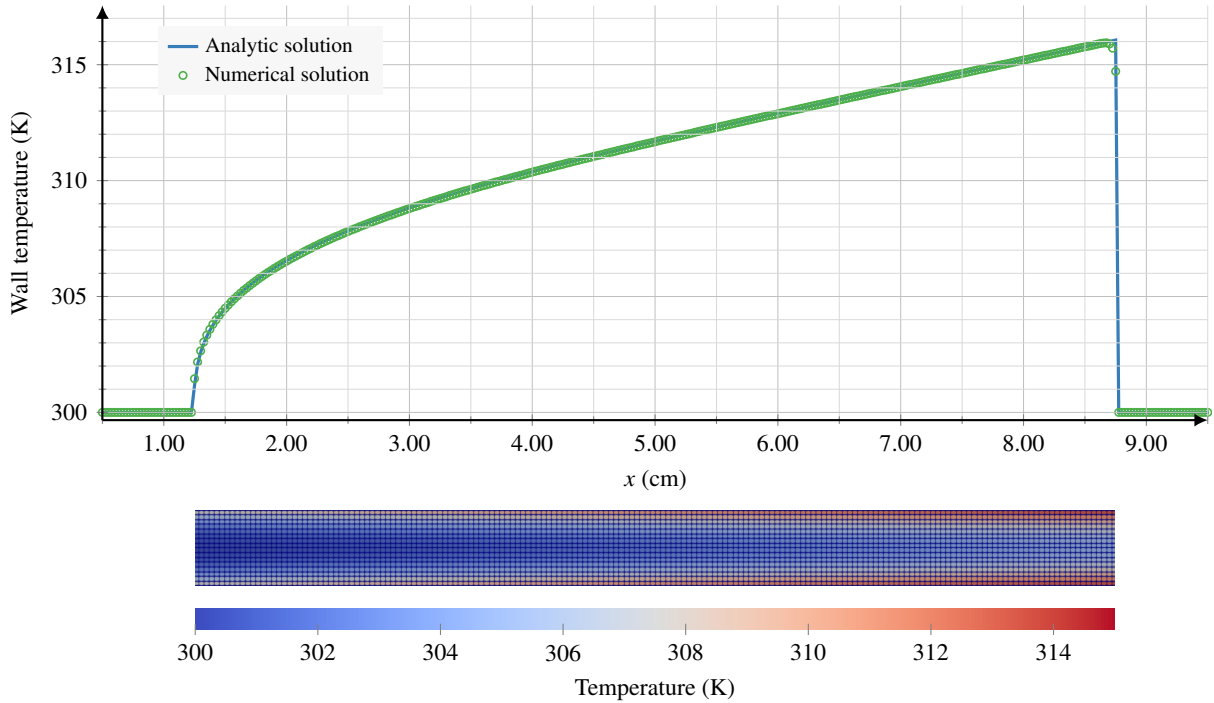
\begin{figure}
  \centering
  \tikzsetnextfilename{wall-temperature}
  \begin{tikzpicture}
    \begin{axis}[
        height=0.40\textwidth,
        width=0.90\textwidth,
        xlabel = $x$ (\si{\centi\meter}),
        ylabel = Wall temperature (\si{\kelvin}),
        xmin = 1.25,
        xmax = 8.75,
        scaled y ticks = false,
        enlargelimits = true,
        x tick label style={
          /pgf/number format/.cd,
          fixed,
          fixed zerofill,
          precision=2,
          /tikz/.cd
        },
        y tick label style={
          /pgf/number format/.cd,
          fixed,
          precision=0,
          /tikz/.cd
        },
        legend style={
          at={(0.05, 0.95)},
          anchor=north west,
        },
        grid = both,
        minor x tick num=1,
        minor y tick num=4,
        table/x expr = 100.0*\thisrow{x},
        table/col sep = space,
        colormap={coolwarm}{
          rgb255=(59,76,192)
          rgb255=(68,90,204)
          rgb255=(77,104,215)
          rgb255=(87,117,225)
          rgb255=(98,130,234)
          rgb255=(108,142,241)
          rgb255=(119,154,247)
          rgb255=(130,165,251)
          rgb255=(141,176,254)
          rgb255=(152,185,255)
          rgb255=(163,194,255)
          rgb255=(174,201,253)
          rgb255=(184,208,249)
          rgb255=(194,213,244)
          rgb255=(204,217,238)
          rgb255=(213,219,230)
          rgb255=(221,221,221)
          rgb255=(229,216,209)
          rgb255=(236,211,197)
          rgb255=(241,204,185)
          rgb255=(245,196,173)
          rgb255=(247,187,160)
          rgb255=(247,177,148)
          rgb255=(247,166,135)
          rgb255=(244,154,123)
          rgb255=(241,141,111)
          rgb255=(236,127,99)
          rgb255=(229,112,88)
          rgb255=(222,96,77)
          rgb255=(213,80,66)
          rgb255=(203,62,56)
          rgb255=(192,40,47)
          rgb255=(180,4,38)
        },
        colorbar horizontal,
        colorbar style={
          xlabel=Temperature (\si{\kelvin}),
          x axis line style={draw=none},
          y axis line style={draw=none},
          point meta min=300,
          point meta max=315,
          xtick={300, 302, ..., 314},
          at={
            (0.084, -0.4),
            anchor=south west,
          },
          width=11.00cm,
          height=0.25cm,
        },
      ]

      \addplot[Set1-B] table[y = T2] {data/wallTemperature.dat};
      \addplot[Set1-C, semithick, only marks, mark=o, mark options={scale=0.7}]
        table[y = T1] {data/wallTemperature.dat};

      \addlegendentry{Analytic solution}
      \addlegendentry{Numerical solution}

      \coordinate (insetPosition) at (axis cs:1.25, 292.5);
    \end{axis}

    \begin{pgfonlayer}{axis ticks}
      \node[above right, inner sep=0] at (insetPosition)
        {\includegraphics[height=1cm,
          width=12.10cm]{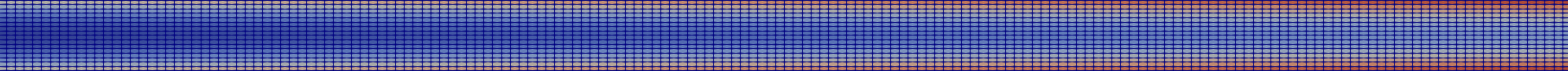}};
      \path (insetPosition) -- +(0,-1.6);
    \end{pgfonlayer}
  \end{tikzpicture}
  \caption{Top: Comparison of analytic (blue) and numerical (green) solution to the extended Graetz problem. Bottom: Temperature contour plot inside a channel, showing the smooth emergence of a thermal boundary layer. }
  \label{fig:wall-temperature}
\end{figure}

The detailed block computes the correct downstream pressure profile to approximately \SI{1}{Pa} for a downstream grid resolution of $\Delta x = \SI{2.5e-1}{\milli\meter}$ and channel resolution $\Delta z = \SI{1.25e-1}{\milli\meter}$, cf.~\cref{fig:graetz-error-norms-1}.
At this resolution, the L2 error norm for the wall temperature profile is well converged as shown in \cref{fig:graetz-error-norms-2}.
\cref{fig:wall-temperature} (left) shows temperature contours inside a channel at steady state in relation to the mesh discretization.
Smooth development of the thermal boundary layer can be observed.
\cref{fig:wall-temperature} (right) provides the corresponding comparison of analyical and computed wall temperature as a function of channel-downstream coordinate. The maximum difference between computed and analytical solution for the wall temperature is $\SI{0.2}{\kelvin}$.
As the purpose of this work is that detailed simulations can be reproduced by the porous approximation, we consider the detailed block resolved at this resolution.
A future experimental validation study of the detailed simulations may be performed at increased near-wall resolution, similar to \cite{skillen2012, torriano2012}, which is outside the scope of this contribution.

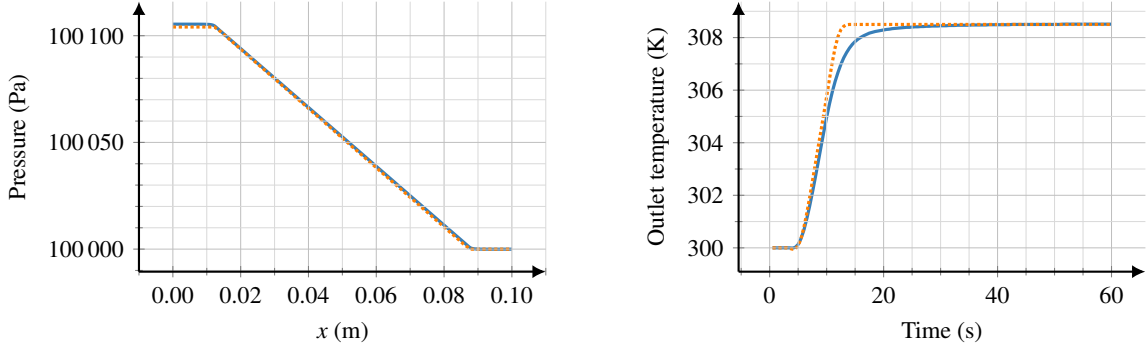
\begin{figure}[tpb]
  \centering
  \begin{subfigure}[b]{0.49\textwidth}
    \centering
    \tikzsetnextfilename{graetz-compare-pressure}
    \begin{tikzpicture}
      \begin{axis}[
          xlabel = $x$ (\si{\meter}),
          ylabel = Pressure (\si{\pascal}),
          table/x = x,
          table/y = p,
          scaled y ticks = false,
          enlargelimits = true,
          x tick label style={
            /pgf/number format/.cd,
            fixed,
            fixed zerofill,
            precision=2,
            /tikz/.cd
          },
          y tick label style={
            /pgf/number format/.cd,
            fixed,
            precision=0,
            /tikz/.cd
          },
          grid = both,
          minor x tick num=1,
          minor y tick num=4,
        ]

        \addplot[p1, c21] table {data/graetz-compare-pressure-detailed.csv};
        \addplot[p2, c22] table {data/graetz-compare-pressure-porous.csv};
      \end{axis}
    \end{tikzpicture}
    \caption{Downstream pressure profiles.}
    \label{fig:graetz-compare-pressure}
  \end{subfigure}
  \begin{subfigure}[b]{0.49\textwidth}
    \centering
    \tikzsetnextfilename{graetz-compare-temperature}
    \begin{tikzpicture}
      \begin{axis}[
          xlabel = Time (\si{\second}),
          ylabel = Outlet temperature (\si{\kelvin}),
          table/x = t,
          enlargelimits = true,
          grid = both,
          minor x tick num=3,
          minor y tick num=1,
        ]

        \addplot[p1, c21] table [y=detailed] {data/graetz-compare-temperature.csv};
        \addplot[p2, c22] table [y=porous]   {data/graetz-compare-temperature.csv};
      \end{axis}
    \end{tikzpicture}
    \caption{Time trace of the average outlet temperature.}
    \label{fig:graetz-compare-temperature}
  \end{subfigure}
  \caption{Comparison of (a) downstream pressure profiles, and (b) average outlet temperature for the PM (dotted orange lines) and DM (solid blue lines) under normal inflow.}
  \label{fig:graetz-comparisons}
\end{figure}

Results for the detailed block at downstream grid resolution of $\Delta x = \SI{2.5e-1}{\milli\meter}$ and channel resolution $\Delta z = \SI{1.25e-1}{\milli\meter}$ are compared with a porous block of $\Delta x = \Delta z = \SI{2}{\milli\meter}$.
The detailed block then consists of 88,000 cells whilst the porous block is constructed of 2,500 cells (\SI{97.2}{\percent} reduction).
\cref{fig:graetz-compare-pressure} shows that the porous approximation matches the downstream pressure profile very closely, while \cref{fig:graetz-compare-temperature} presents the time evolution of the average outlet temperature.
The porous block matches the steady-state outlet temperature of the detailed block, by construction of the source term.
In this simple case, a surprisingly close agreement in terms of transient temperature evolution can also be observed, with small deviations around \SI{1}{\kelvin} at the maximum.

\subsection{Turning flow}
\label{subsec:turning-flow}
With the analytical comparisons in place from the previous subsection, we now consider a more realistic geometry and flow pattern:
turning flow induced by pass washers mimicked by inlet at bottom left leg and outlet at the top right leg.
Gravity is no longer neglected.
We consider both an idealized oil with constant thermophysical properties as listed in \cref{subsec:analytical-comparison}, as well as the MIDEL 7131 where the temperature-dependent expressions are given in \cref{sub:theory_equations}.
For the geometry, we use a left and right leg spacing of $l = \SI{7}{\milli\meter}$, channel gap $h_c = \SI{4}{\milli\meter}$, disk height $h_p = \SI{15}{\milli\meter}$, disk length $L = \SI{51}{\milli\meter}$, with $N = 19$ disks, and consequently 20 channels in a single pass.
All channels are of the same width, $h_c$, for the remainder of this contribution.
These dimensions correspond roughly to the CIGRE cases~\citep{cigre2016}.
They have also been studied in \citep{skillen2012,torriano2012}, though with different thermophysical oil properties.

At the inlet, we again apply the same prescribed mass-flow rate, $\dot{m} = \SI{2.167e-3}{\kilogram\per\second}$.
No-slip boundary conditions are specified on the walls, and a constant heat-flux $q = \SI{2336.4}{\watt\per\meter\squared}$ is applied on the sides of the winding turns, i.e.~the channel walls.
The choice of heat-flux is motivated by characteristic values from \citep{cigre2016,skillen2012,torriano2012}.
Temperature is kept at its fixed initial value of $\SI{300}{\kelvin}$ at the inlet.
The pressure is fixed at $\SI{1e5}{\pascal}$ at the outlet, with a zero-value Neumann boundary condition at the inlet.
The boundary conditions consequently resemble a realistic transformer winding flow setting.

The top-oil temperature is a measure that can be used to compare the DM and PM.
It can be calculated as the time trace of the mass-averaged temperature at the outlet patch,
\begin{equation}
  T_\text{top} =
    \frac{\int \rho c_p T \vct{u} \cdot \mathrm{d} \vct{A} (t)}%
         {\int \rho c_p \vct{u} \cdot \mathrm{d} \vct{A}}.
\end{equation}
For constant oil properties, the top-oil temperature can be calculated directly from energy conservation and the integrated power supplied to the fluid, $Q =  \dot m c_p \Delta T$.
In particular, we find a top-oil temperature of $T_\mathrm{top} = T (t=0) + \Delta T \approx \SI{314.2}{\kelvin}$.

\begin{figure}[tpb]
  \centering
  \begin{subfigure}[b]{0.49\textwidth}
    \centering
    \tikzsetnextfilename{turning-flow-constant}
    \begin{tikzpicture}
      \begin{axis}[
          xlabel = Time (\si{\second}),
          ylabel = Top oil temperature (\si{\kelvin}),
          xmax = 200,
          enlargelimits = true,
          grid = both,
          minor x tick num=1,
          minor y tick num=1,
          table/x = t,
          restrict x to domain = 0:200,
        ]

        \addplot[p1, c21] table [y=d1] {data/turning-flow.csv};
        \addplot[p2, c22] table [y=p1] {data/turning-flow.csv};
      \end{axis}
    \end{tikzpicture}
    \caption{Constant oil properties.}
    \label{fig:turning-flow-comparison-constant}
  \end{subfigure}
  \begin{subfigure}[b]{0.49\textwidth}
    \centering
    \tikzsetnextfilename{turning-flow-dependent}
    \begin{tikzpicture}
      \begin{axis}[
          xlabel = Time (\si{\second}),
          ylabel = Top oil temperature (\si{\kelvin}),
          xmax = 300,
          enlargelimits = true,
          grid = both,
          minor x tick num=1,
          minor y tick num=1,
          table/x = t,
          restrict x to domain = 0:302,
        ]

        \addplot[p1, c21] table [y=d2] {data/turning-flow.csv};
        \addplot[p2, c22] table [y=p2] {data/turning-flow.csv};
      \end{axis}
    \end{tikzpicture}
    \caption{Temperature-dependent oil properties.}
    \label{fig:turning-flow-comparison-dependent}
  \end{subfigure}
  \caption{Time traces of the mass-averaged top-oil temperature for the PM (dotted orange lines) and the DM (solid blue lines).}
  \label{fig:turning-flow-comparison}
\end{figure}
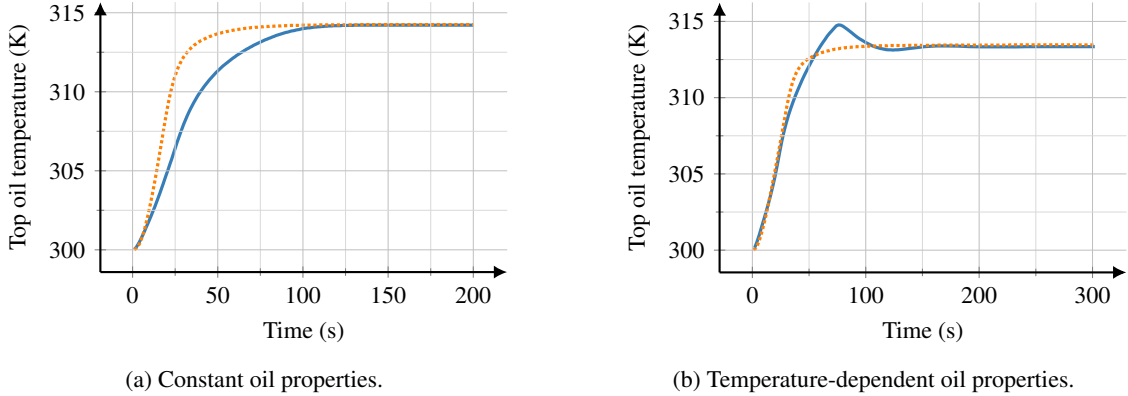

\Cref{fig:turning-flow-comparison} shows a comparison of the top-oil temperatures computed by the DM and the PM for both constant (a) and temperature-dependent (b) oil properties.
In both figures, there is a difference between the topoil temperature for the porous and resolved simulation during the transient evolution.
This is as expected, since the resulting flow pattern is different by construction: in the porous block, heat is transferred by a plug-flow velocity profile, which \emph{on average} yields the same throughput as in the resolved geometry, where the velocity profile is of parabolic shape inside the channels.
However, as the flows tend towards steady state, a good agreement between the traces is observed.
Specifically, the calculated outlet heat flux for constant oil properties agrees well for the detailed and porous pass, respectively.
With temperature-dependent oil properties there are nonlinear effects which give rise to small-scale temperature fluctuations, such as hot-plumes.
These are captured by the DM, but absent in the PM.
Nonetheless, the maximum deviation between the DM and PM is \SI{1.5}{\kelvin} in the transient phase, while in steady state, the top-oil temperatures agree.
Further, the outlet heat fluxes align.
We stress that a mere \num{5040} cells are necessary to produce similar outlet heat fluxes to the detailed pass at \num{212320} cells resolution --- a \SI{97.7}{\percent} reduction.

\section{Demonstration}
\label{sec:demonstration}
In the previous sections, we have illustrated that the method is capable of reproducing analytical wall temperatures in simple laminar channel flow in fully resolved simulations.
The porous approximation to a number of micro-channels stacked atop each other can equally well provide the averaged pressure drop, mass flux, and heat flux over the domain.
We have shown that the transient and steady-state top-oil temperature measured at the outlet agrees well for a fully resolved pass and its porous approximation.
Further, temperature-dependent fluid properties have been introduced, and we have indicated that the porous approximation holds to a reliable degree also in this case.

Building on that, we here construct \emph{stacks} of detailed and porous blocks to show the significant problem-size reduction that can be obtained by the method presented in this work.
The output from coupled porous passes is used to deliver boundary conditions to a fully resolved pass.
\cref{subsec:2D-demo} considers 2D simulations of a four-pass winding, qualitatively corresponding to series of flow transitions as discussed in \cref{subsec:turning-flow}.
In \cref{subsec:3D-demo}, we present a full 3D numerical simulation of a whole transformer leg consisting of 72 passes.
The transformer leg measures 4 passes in vertical direction, with the turns of the winding azimuthally separated by 18 block washers.

\subsection{2D four pass winding}
\label{subsec:2D-demo}
\begin{figure}
  \centering
  \includegraphics[width=15cm]{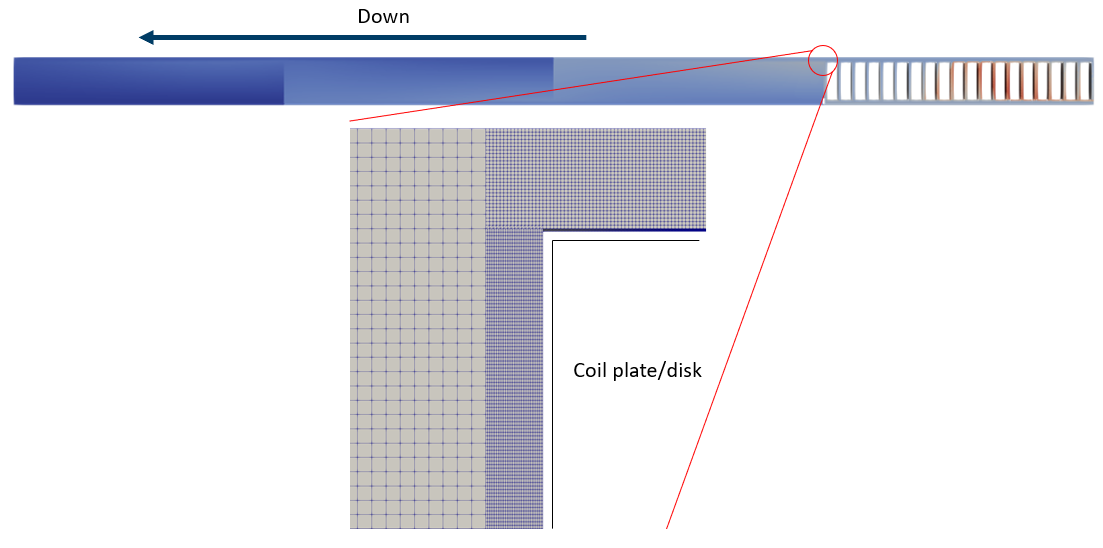}
  \caption{Mesh details of the 2D joint DM-PM computational domain.
    The top horizontal column depicts the four-pass arrangement (bottom to the left), with the color indicating typical temperature fluctuations (blue-cold, red-warm).
  The coarse cells in the mesh zoom are placed in the porous domain.}
  \label{fig:2D-detail}
\end{figure}

A \SI{1.460}{\meter} transformer winding that consists of 4 passes and 80 channels between turns is computed fully resolved by four detailed blocks.
This detailed stack then consists of 849280 cells with a resolution of \SI{0.25}{\milli\meter} per cell in the $x$ direction and \SI{0.25}{\milli\meter} and \SI{0.125}{\milli\meter} per cell in the $z$ direction for the legs and channels, respectively.
The same winding is also approximated by three porous passes with a fully resolved pass on top.
\Cref{fig:2D-detail} illustrates changes in the mesh as the computational model switches between the porous approximation (bottom) and the detailed model (top).
With a porous resolution of 1 cell per \si{\milli\meter}, and identical detailed resolution, the approximated winding then consists of 283495 cells.
This gives a cell reduction of \SI{67}{\percent}.
Note that the single detailed pass accounts for \SI{75}{\percent} of the necessary cells.
Both simulations employ the same numerical schemes and solution criteria.
Boundary conditions are those of \cref{subsec:turning-flow}.

In terms of computational run times, the detailed model computes 1--3 timesteps per physical second when the code runs on 2 nodes with dual-socket Intel Xeon 4116 processors, corresponding to 48 physical cores in total.
The porous model achieves the same run-time performance when running on a single node with 2 physical cores, a reduction in the computing-power requirement of \SI{95}{\percent}.
Clearly, the porous model outperforms the detailed model in terms of run time and resource allocation.

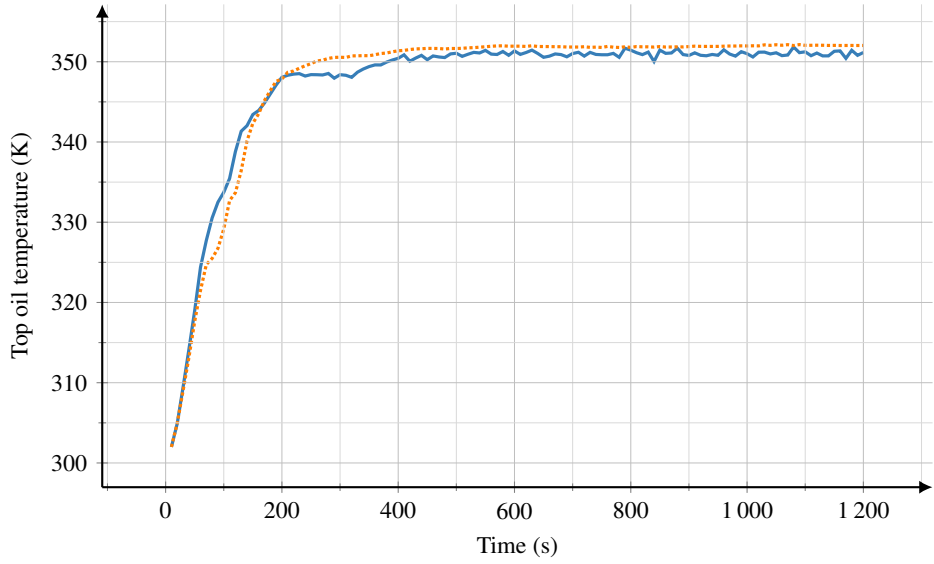
\begin{figure}[tpb]
  \centering
  \tikzsetnextfilename{heat-flux-stack}
  \begin{tikzpicture}
    \begin{axis}[
      height = 0.45\textwidth,
      width = 0.70\textwidth,
      xlabel = Time (\si{\second}),
      ylabel = Top oil temperature (\si{\kelvin}),
      enlargelimits = true,
      grid = both,
      minor x tick num=1,
      minor y tick num=1,
      table/x = t,
      ]

      \addplot[p1, c21] table [y=detailed] {data/heat-flux-stack.csv};
      \addplot[p2, c22] table [y=porous]   {data/heat-flux-stack.csv};
    \end{axis}
  \end{tikzpicture}
  \caption{Mass averaged top-oil temperature for the PAM (dotted orange line) and the DM (solid blue line). Small fluctuations seen for the DM are a result from averaging over the thermal fluctuations in the unsteady flow exiting the transformer. With the PAM, these fluctuations are not resolved, and the top-oil temperature is smoother.}
  \label{fig:heat-flux-stack}
\end{figure}

Cf.~\cref{fig:heat-flux-stack}, which shows that deviations are identifiable throughout the transient evolution and most pronounced close to the onset of the statistical steady state for the detailed stack.
The maximum discrepancy in top-oil temperature amounts to \SI{4}{\kelvin} (relative error \SI{1.2}{\percent}) in the transient phase, with the steady state values overlapping to within $\SI{1}{\kelvin}$ (relative error \SI{0.3}{\percent}).
Referring to \cref{fig:turning-flow-comparison-dependent}, we note that this is likely due to the transient nature of the hot-streaks, which are absent in the porous part of the porous-approximate model.
Precisely, as discussed in \cref{subsec:turning-flow}, the velocity profile through the porous domain is \emph{locally} different from the combined effect of parabolic profiles from the channels.
The time averaged heat flux in statistical steady state, is accurately reproduced in the porous-approximate model.
The goal of the approximate model is to provide similar quantities for top-oil and averaged-disk temperature, compared to a fully resolved stack.
\Cref{fig:heat-flux-stack} shows the evolution of the top-oil temperature (outlet averaged temperature).
Clearly, the temperature field is more inhomogeneous in the DM, with the summed effect of the hot-streaks from each pass appearing through ``bursts'' in top-oil temperature.
\cref{fig:heat-flux-stack} shows the mass-averaged top-oil temperature, with temperature inhomogeneity further pronounced through the temperature dependence of the density, resulting in fluctuations.
In an experimental setup, usual measurements of top oil temperature do not show these fluctuations, due to the finite thermal response time of the temperature sensors.
As the details of the hot-streaks are absent in most of the PAM, a smoother temperature time-trace is observed.
Overall, and particularly in steady-state, the agreement between the models is promising.

\begin{figure}[tbp]
  \centering
  \tikzsetnextfilename{compare-plate-temperatures}
  \begin{tikzpicture}
    \begin{axis}[
        height = 0.45\textwidth,
        width = 0.70\textwidth,
        xlabel = Plate number,
        ylabel = Average plate temperature (\si{\kelvin}),
        enlargelimits = true,
        grid = both,
        minor x tick num=1,
        minor y tick num=1,
        table/x = n,
      ]

      \addplot[densely dotted, c21, dp1] table [y=dm]  {data/compare-plate-temperatures.csv};
      \addplot[densely dotted, c22, dp2] table [y=pam] {data/compare-plate-temperatures.csv};
    \end{axis}
  \end{tikzpicture}
  \caption{Contour-integrated plate temperature in the top pass at $t = \SI{1200}{\second}$ for the DM (blue circles) and the PAM (orange stars).
  Constant time step of $\Delta t = \SI{1e-3}{\second}$.}
  \label{fig:compare-plate-temperatures}
\end{figure}
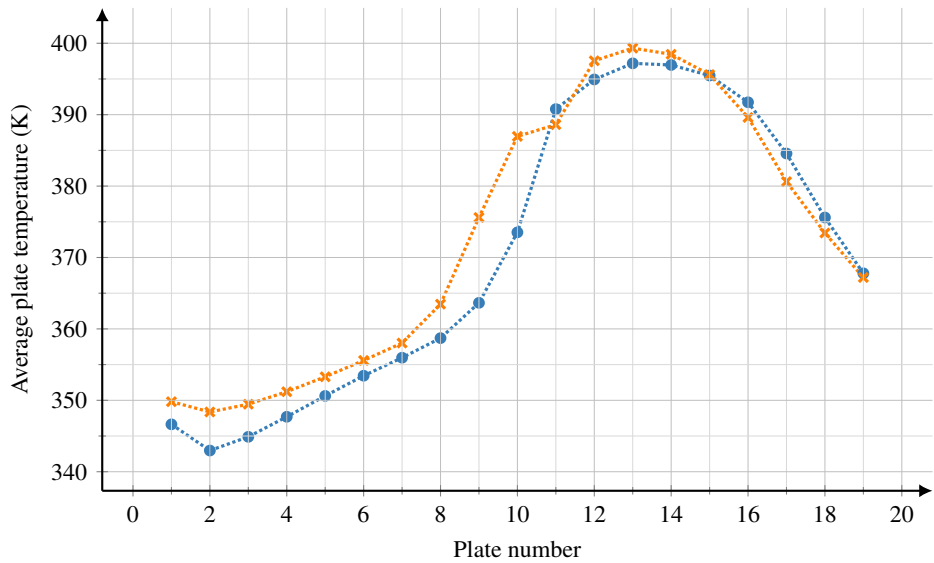

In \cref{fig:compare-plate-temperatures}, we compare the average plate temperatures for the top pass in both models at $t = \SI{1200}{\second}$.
To this end, the temperature of plate-adjacent cells is sampled along four line segments, constituting a closed contour around each plate.
The relation among cell-centre and face-value temperature is provided via
\begin{equation}
  T_\mathrm{face} = T_\mathrm{cell} + \frac{1}{2} \Delta \grad T|_\mathrm{face},
\end{equation}
with the face-to-centre distance $\Delta$ and the fixed temperature-gradient boundary condition at the plate walls entering via $\grad T|_\mathrm{face}$.
The contour-averaged plate temperature is then found by integration of the face temperature along the sample.
From \cref{fig:compare-plate-temperatures}, it is evident that the flow pattern in the top pass differs predominantly around the pass centre.
Deviations in plate temperature vary between \SI{1}{\kelvin} (plate 13) and \SI{11}{\kelvin} (plate 9).
We note that the hot-spot location is accurately predicted by the porous-approximated model, with a problem size reduction of \SI{67}{\percent}.
The observed discrepancies in flow pattern and resulting plate temperature distribution are attributed to the sensitivity of these quantities with respect to the porous-detailed transition temperature profile.
This is similar to the findings of \citet{skillen2012}, who have found flow-pattern sensitivity with respect to inlet temperature profile in their simulations.

\subsection{3D four pass transformer leg}
\label{subsec:3D-demo}
\begin{figure}
  \centering
  \includegraphics[width=16cm]{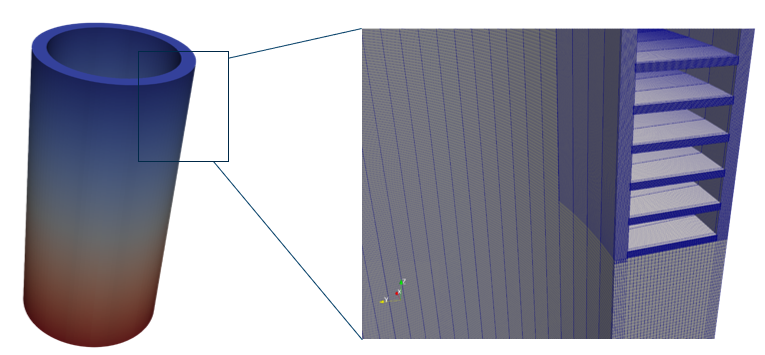}
  \caption{Mesh details of the 3D joint DM-PM computational domain.
    The 3D simulation domain corresponds to an entire leg, as shown in the leftern part of the figure.
    Colors correspond to the pressure, with red denotion high values, and blue low values.
  The right part of the figure shows the transition between the part modelled by the PM, and the details associated with the DM.}
  \label{fig:3D-detail}
\end{figure}

Full 3D simulations are immensely costly for fully resolved flow calculations: 72 passes of resolution as described in \cref{subsec:2D-demo} account for at least \num{15 000 000} cells.
Following the modelling strategy outlined in this work, a reasonable assessment of selected top passes can be achieved at significantly reduced computational cost.
Clearly, the number of detailed top passes is dependent on the specific design.
In \cref{subsec:2D-demo}, we have shown that the approximate model can confidently reproduce hot-spot location and corresponding plate temperature, as well as top-oil temperature and top oil heat-flux, as compared to the fully resolved stack.
Here, we illustrate the model's three-dimensional capability by considering top passes at two distinct azimuthal locations.
In total, 18 azimuthal stacks consisting of 4 passes each are considered.
This represents the transformer leg geometry in \cite{cigre2016}, taking into account the azimuthal block washers/sticks.
\Cref{fig:3D-detail} illustrates a resolved block with the computational domain adapted to the physical presence of plates, immersed in surrounding low-resolution porous blocks, representing the majority of the transformer leg.

As this section mainly serves illustrative purposes, we reduce the resolution to \SI{0.5}{\milli\meter} in $x$ direction and \SI{0.25}{\milli\meter} in $z$ direction, and a resolution of 1 cell per 3 degrees in the azimuthal direction.
\Cref{fig:graetz-error-norms} (blue dots) show that this resolution should capture larger scale phenomena, but at less accuracy than the more resolved results presented in \cref{subsec:2D-demo}.
The total number of cells for a transformer leg resolved at this level then amounts to \num{777400} cells
This is approximately \SI{10}{\percent} that of a comparable fully resolved leg and \SI{5}{\percent} of a fully resolved leg at the resolution employed in \cref{subsec:2D-demo}.
In the following, we employ the steady-state permeability given by \cref{eq:radial-permeability}.
The double-infinite sum for $h = \SI{4}{\milli\meter}$, $H=\SI{365}{\milli\meter}$, $N=20$, $R_1 = \SI{307}{\milli\meter}$,  $R_2 = \SI{358}{\milli\meter}$, and $\Delta \theta = \pi / 9$ converges to
\begin{equation}
  \kappa \approx \SI{5.66e-5}{\meter\squared}.
\end{equation}

\begin{figure}
  \centering
  \tikzsetnextfilename{heat-distribution}
  \begin{tikzpicture}
    \begin{polaraxis}[
        height = 0.45\textwidth,
        width = 0.45\textwidth,
        set layers = default,
        enlargelimits = true,
        axis line style = {thick, -},
        xtick distance = 45,
        xticklabel shift = 4pt,
        xticklabel = $\pgfmathprintnumber{\tick}^\circ$,
        ytick pos = both,
        ytick = {0, 0.2, ..., 1.0},
        yticklabel style = {above},
        y tick label style={
          /pgf/number format/.cd,
          fixed,
          fixed zerofill,
          precision=1,
          /tikz/.cd
        },
        grid = both,
        minor x tick num = 1,
        scatter/classes = {
          style1={solid, /pgfplots/c31, /pgfplots/dp1},
          style2={solid, /pgfplots/c32, /pgfplots/dp2},
          style3={solid, /pgfplots/c33, /pgfplots/dp3}
        },
      ]

      \addplot[scatter, scatter src=explicit symbolic, densely dotted, c21, data cs=polarrad]
        table[meta=style] {data/heat-distribution.csv} -- cycle;

      \draw[very thick, /pgfplots/c32] (40, 0.5853571667841480641) circle(5pt);
      \draw[very thick, /pgfplots/c33] (280, 0.9703009328914133924) circle(5pt);
    \end{polaraxis}
  \end{tikzpicture}
  \caption{Polar plot of normalized azimuthal power distribution.
  The resolved passes are inserted at $\theta = 40^\circ$ and $\theta = 280^\circ$, indicated by an orange star and a green triangle, respectively.}
  \label{fig:heat-distribution}
\end{figure}
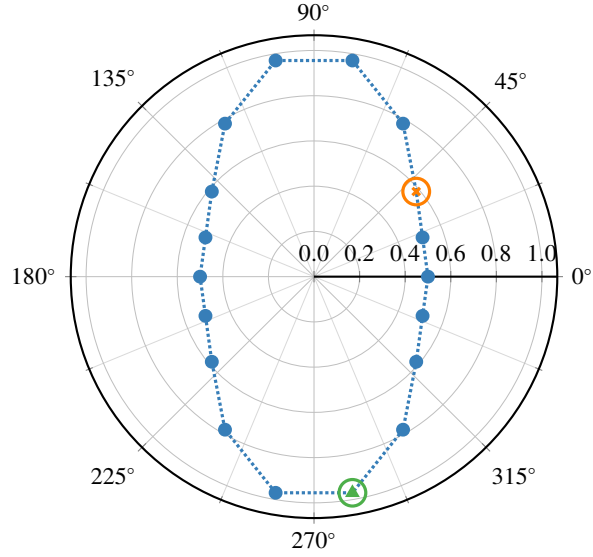

To demonstrate the model's capability in terms of non-uniform azimuthal heat generation, we distribute the heat flux and heat source from \cref{subsec:2D-demo} azimuthally according to
\begin{equation}
  \label{eq:power-distribution}
  P = P_0 \left[\sin^2 \theta + \frac{1}{2} \cos^4 \theta \right].
\end{equation}
This results in a normalized power distribution as shown in \cref{fig:heat-distribution}.
Recall that the supplied heat enters the model formulation as heat-flux boundary conditions in the DM and as a volumetric heat-source in the PAM/PM.
Consequently, the heat flux and heat source are modulated by $P/P_0$ described in \cref{eq:power-distribution}.
$P_0$ denotes the respective power for the cases considered in \cref{subsec:2D-demo}.
The mass-flow rate from the 2D case is scaled by the 3D inlet area.
We consider both uniform heating at each azimuthal stack (UH) and nonuniform, azimuthally distributed heating (NH) according to \cref{eq:power-distribution}.
The UH and the NH cases resemble the 2D PAM stack from \cref{subsec:2D-demo} at the azimuthal sections, were top passes are fully resolved, i.e. at $\theta = 40^\circ$ and $\theta = 280^\circ$.
The remainder of the leg is modelled fully porous.
The dissipated heat in the NH case, found from averaging \cref{eq:power-distribution}, is on average \SI{31.5}{\percent} less than the UH case.

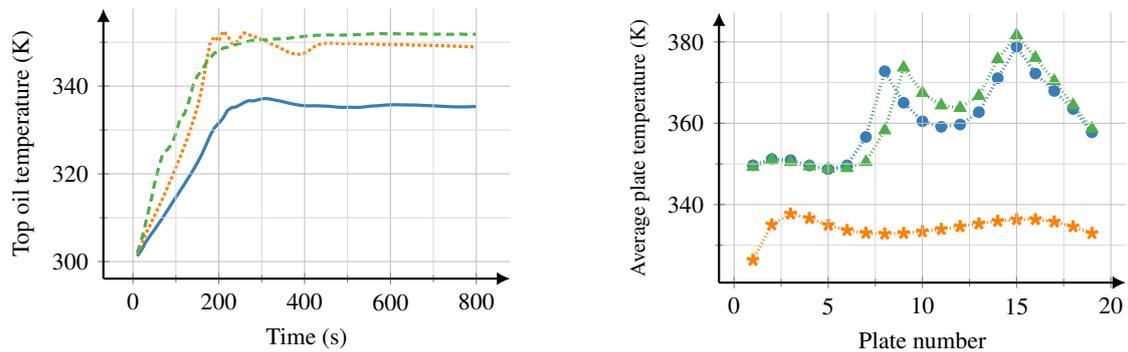
\begin{figure}
  \centering
  \begin{subfigure}[t]{0.49\textwidth}
    \centering
    \tikzsetnextfilename{3d-results-topoil}
    \begin{tikzpicture}
      \begin{axis}[
          xlabel = Time (\si{\second}),
          ylabel = Top oil temperature (\si{\kelvin}),
          enlargelimits = true,
          grid = both,
          minor x tick num=1,
          minor y tick num=1,
          table/x = t,
          restrict x to domain = 0:800,
        ]

        \addplot[p1, c31] table [y=nonunif] {data/3d-results-topoil.csv};
        \addplot[p2, c32] table [y=unif]    {data/3d-results-topoil.csv};
        \addplot[p3, c33] table [y=2dpam]   {data/3d-results-topoil.csv};
      \end{axis}
    \end{tikzpicture}
    \captionsetup{margin={3pt,5pt}}
    \caption{Mass averaged top-oil temperature for the azimuthally uniform (dotted orange line) and nonuniform (solid blue line) heating, compared with the 2D PAM (dashed green line).}
    \label{fig:3d-results-topoil}
  \end{subfigure}
  \begin{subfigure}[t]{0.49\textwidth}
    \centering
    \tikzsetnextfilename{3d-results-average}
    \begin{tikzpicture}
      \begin{axis}[
          xlabel = Plate number,
          ylabel = {\footnotesize Average plate temperature (\si{\kelvin})},
          enlargelimits = true,
          grid = both,
          minor x tick num=1,
          minor y tick num=1,
          table/x = n,
          thin,
          densely dotted,
        ]

        \addplot[dp1, c31] table [y=uniform] {data/3d-results-average.csv};
        \addplot[dp2, c32] table [y=theta1]  {data/3d-results-average.csv};
        \addplot[dp3, c33] table [y=theta2]  {data/3d-results-average.csv};
      \end{axis}
    \end{tikzpicture}
    \captionsetup{margin={5pt,3pt}}
    \caption{Contour-integrated plate temperature in the top pass at $t = \SI{800}{\second}$.
      Blue circles denote uniform heating.
    Orange stars lable non-uniform heating at $\theta = 40^\circ$, and green triangles at $\theta = 280^\circ$, respectively.}
    \label{fig:3d-results-average}
  \end{subfigure}
  \caption{(a) Top-oil temperature time traces and (b) average plate temperature derived from 3D simulations with uniform and nonuniform heating.}
  \label{fig:3d-results}
\end{figure}

\Cref{fig:3d-results-topoil} shows a comparison of the top-oil temperature for the 2D PAM, UH and NH case.
The temperature rise in the NH case is approximately \SI{29}{\percent} less than for the UH, in agreement with the averaged dissipated heat being less in the NH case.
Comparing the 2D PAM and UH top-oil temperature, we observe agreement to within \SI{0.8}{\percent} (approximately \SI{2.8}{\kelvin}) in steady state.
It should be noted that the ratio of dissipated heat to fluid volume from the 2D PAM case of \cref{subsec:2D-demo} is approximately equal to the UH 3D setting.
Deviations are likely due to the average calculation across the entire outlet, which consists of 2 resolved top passes, and 16 porous passes in the UH case.

\Cref{fig:3d-results-average} depicts the contour-integrated plate temperatures in the top pass at $t = \SI{800}{\second}$ for the UH and NH cases.
Contours are taken along the plates in the azimuthal planes at $\theta = 40^\circ$ and $\theta = 280^\circ$, respectively.
The UH case shows two distinct regions of enhanced temperature, with a first peak at plate 8 and a second, higher peak in the vicinity of plate 15.
In the NH case at $\theta = 280^\circ$, corresponding to $P \approx 0.97 P_0$, two hot-spots are observed close to plate 9 and plate 15.
At $\theta = 40^\circ$, corresponding to $P \approx 0.59 P_0$, the highest temperature is measured near plate 3, with the temperature profile significantly changed.
Contrasting \cref{fig:3d-results-average} and \cref{fig:compare-plate-temperatures}, we note that the 2D profiles are elevated in temperature and feature only one hot-spot, located at plate 13.
Naturally, the three-dimensional flow patterns are different, most notably due to the no-slip boundary condition on the velocity at the azimuthal block washers ("sticks").

\begin{figure}
  \centering
  \includegraphics[width=\linewidth]{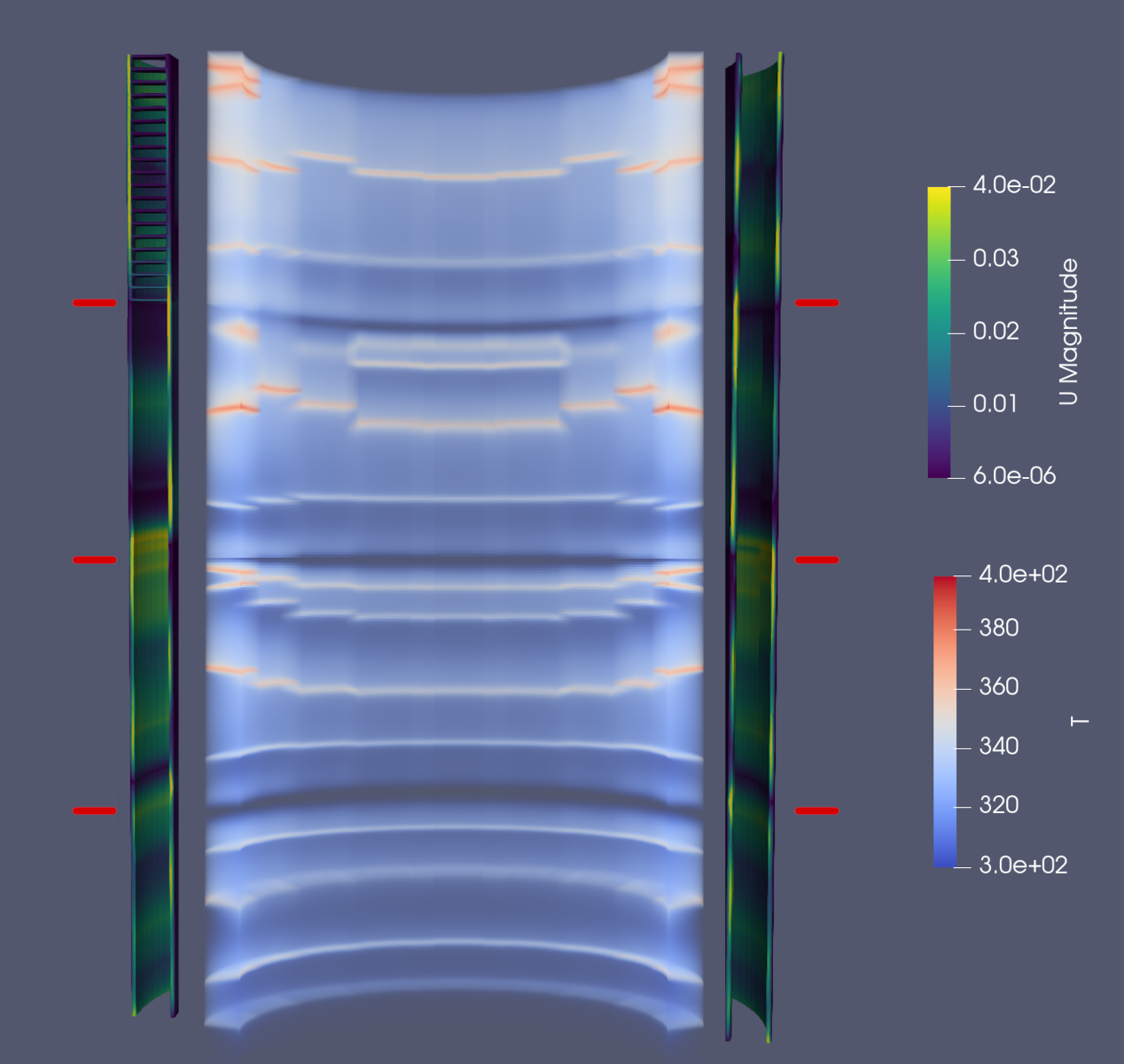}
  \caption{Volume render of the temperature field (red/blue) in a half section of the transformer, with velocity magnitude (yellow/green) at the cut plane shown at the sides. The red lines indicate the blocks separating the four passes in the vertical direction. The distinct jumps in hotspot positions are due to the  nonuniform heating load combined with the azimuthal block washers that separate the flow into 9 distinct regions per 180$^\circ$ azimuthally.}
  \label{fig:3d-results-volT}
\end{figure}

\Cref{fig:3d-results-volT} shows a 3D volumetric rendering of the temperature field (red/blue) in a half section of the transformer for the NH case.
The dissipated power (heating) is lowest at the centre and highest at the sides of this plot.
The azimuthal evolution of the hot-spot location can be recognized also here, with discrete jumps due to the azimuthal block washers.
More research is needed to clarify the role of three-dimensional effects on the emergence of the observed double hot-spot at locations with highest heating.

\begin{figure}
  \centering
  \includegraphics[width=0.8\linewidth]{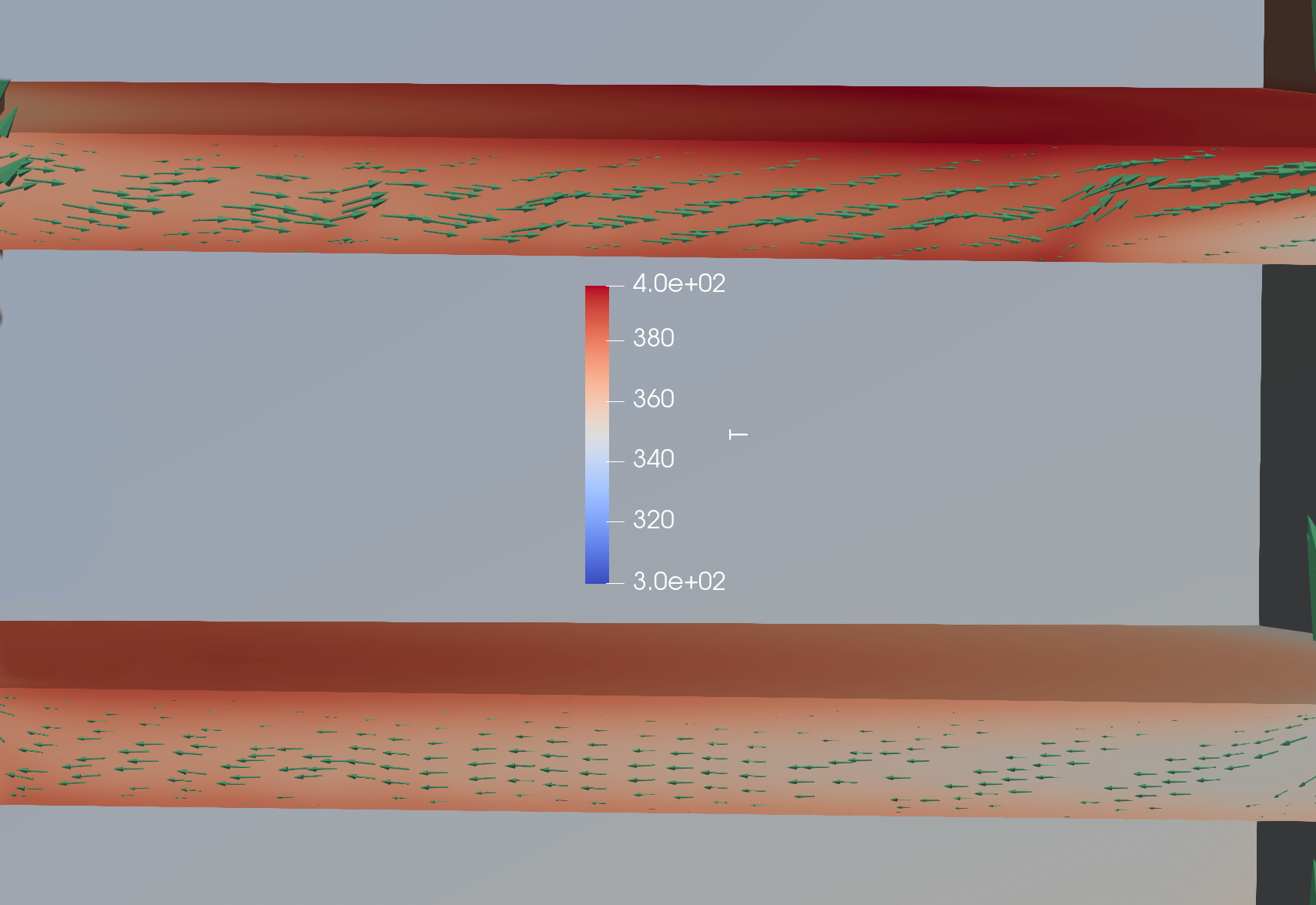}
  \caption{Close-up view of the temperature distribution (blue/red) and velocity field (green vectors, shown every 7th grid point) inside the channel where the hot-spot is located and the channel immediately below it.}
  \label{fig:3d-results-zoom}
\end{figure}

Finally we show in \cref{fig:3d-results-zoom} a close-up view of the hot spot location and the channel below it, in the resolved region of the NH case.
This figure indicates that flow reversal can occur between two channels in the same pass, and that flow patterns driven by thermal convection can be seen in the channels.

\section{Conclusions}
\label{sec:conclusion}
In this work, we have proposed a simulation methodology that combines a porous-medium approach with fully resolved simulations for studying complex flow and heat-transfer problems.
The novel approach has been applied to study oil flow and heat transfer in large electrical power transformers and we have demonstrated a significant reduction in computational cost compared to a fully resolved approach.

The main idea is to approximate the flow through complex geometries as the flow through a porous medium.
In particular, by approximating a typical transformer pass with an anisotropic porous medium, the number of grid cells can be reduced by up to two orders of magnitude when compared to a fully resolved geometry.
We have shown that the porous description retains the important characteristics of the macroscopic quantities of the fully resolved steady-state flow of a transformer pass, in agreement with previous studies.
Consequently, we propose a novel approach in which a transformer winding is modelled as a combination of porous blocks together with full-resolution passes for the regions in which the main phenomena of interest arise.
This approach allows for numerical simulations of an entire transformer leg at considerably reduced computational effort as compared to a fully resolved simulation.

Our approach has been demonstrated for both a 2D and a full 3D representation of a transformer leg.
In the 2D case, the transformer leg consisted of 4 winding passes, where the top pass is fully resolved and the bottom three passes are approximated as porous blocks.
When compared to a fully resolved simulation, we observe that the plate-surface temperature-distribution profile in the top stack is accurately captured despite the reduction in resolution in the bottom passes.
Our approach accurately reproduces both the magnitude and the location of the hot spot located in the top pass.

In 3D, a fully resolved simulation would require an excessive amount of computation resources.
We have demonstrated that our approach allows computations of an entire transformer leg at a significant reduction of the required computational resources.
The 3D simulations with azimuthally uniform heating show similarities with the 2D computations in terms of the top-oil temperature; however, flow patterns and temperature distribution differ significantly.
Going further and considering an azimuthally non-uniform heating, the vertical location of the hot-spot is found to vary in a non-linear fashion, and a double hot-spot emerges in some locations.
Whether the emergence of the double hot-spot and sensitivity of the temperature profile with respect to heat load can be attributed to three-dimensional effects should be investigated further in future work.
In future work, the method will be verified against experimental measurements presently being conducted.

It has been demonstrated that the approach presented here can significantly reduce the computational expense of simulating a full power transformer, while still resolving the key physics.
This will enable larger scale parameter studies on e.g.~coolant properties, geometric features, or forcing flow rates in a transformer.
It also paves the way for developing optimization methods on top of the fluid dynamics models.
One may also envision an extension to the porous approximation that models the resulting temperature profile at the transition based on averaged heat-flux.
Another related avenue is to use the porous-medium solution as a preconditioner to the fully resolved solver.
For highly complex geometries or highly transient cases, this might provide a fruitful approach.

\section*{Acknowledgment}
This work was performed within the project ``Thermal Modelling of Transformers'' (project number: 255178) funded by the Research Council of Norway, Statnett, Hafslund and Lyse Nett.


\end{document}